\documentclass[twocolumn]{revtex4}

\usepackage{amssymb}
\usepackage{graphicx}

\begin{document}

\title{
\textit{q}\textbf{-deformed statistical-mechanical property in the dynamics
of trajectories en route to the Feigenbaum attractor}
}

\author{A. Robledo}
\affiliation{Instituto de F\'{\i}sica, Universidad Nacional Aut\'onoma de M\'exico,
Apartado Postal 20-364, M\'exico 01000, Distrito Federal, M\'exico}

\author{Luis G. Moyano}
\affiliation{Departamento de Matem\'aticas and Grupo Interdisciplinar de Sistemas Complejos,
Universidad Carlos {\rm III} de Madrid, 28911 Legan\'es, Madrid, Spain}

\begin{abstract}
We demonstrate that the dynamics towards and within the Feigenbaum attractor
combine to form a $q$-deformed statistical-mechanical construction. The rate
at which ensemble trajectories converge to the attractor (and to the
repellor) is described by a $q$-entropy obtained from a partition function
generated by summing distances between neighboring positions of the
attractor. The values of the $q$-indices involved are given by the unimodal
map universal constants, while the thermodynamic structure is closely
related to that formerly developed for multifractals. As an essential
component in our demonstration we expose, in great detail, 
the features of the dynamics of trajectories that either evolve
towards the Feigenbaum attractor or are captured by its matching repellor.
The dynamical properties of the family of periodic superstable cycles in
unimodal maps are seen to be key ingredients for the comprehension of the
discrete scale invariance features present at the period-doubling transition
to chaos. Elements in our analysis are the following. (i) The preimages of
the attractor and repellor of each of the supercycles appear entrenched into
a fractal hierarchical structure of increasing complexity as period doubling
develops. (ii) The limiting form of this rank structure results in an
infinite number of families of well-defined phase-space gaps in the
positions of the Feigenbaum attractor or of its repellor. (iii) The gaps in
each of these families can be ordered with decreasing width in accordance with
power laws and are seen to appear sequentially in the dynamics generated by
uniform distributions of initial conditions. (iv) The power law with
log-periodic modulation associated with the rate of approach of trajectories
towards the attractor (and to the repellor) is explained in terms of the
progression of gap formation. (v) The relationship between the law of rate of
convergence to the attractor and the inexhaustible hierarchy feature of the
preimage structure is elucidated. (vi) A ``mean field'' evaluation of the
atypical partition function, a thermodynamic interpretation of the time
evolution process, and a crossover to ordinary exponential statistics are
given. We make clear the dynamical origin of the anomalous thermodynamic
framework existing at the Feigenbaum attractor.



\end{abstract}
\maketitle
\section{Introduction}

A fundamental question in statistical physics is whether the
structure of ordinary equilibrium statistical mechanics falters when its
fundamental properties, phase-space mixing and ergodicity, breakdown. The
chaotic dynamics displayed by dissipative nonlinear systems, even those of
low dimensionality, possesses these two crucial conditions, and acts in
accordance with a formal structure analogous to that of canonical
statistical mechanics, in which thermodynamic concepts meet their dynamical
counterparts \cite{beck1}. At the transition between chaotic and regular
behavior, classically represented by the Feigenbaum attractor \cite{beck1},
the Lyapunov exponent vanishes and chaotic dynamics turns critical.
Trajectories cease to be ergodic and mixing; they retain memory of
their initial positions and fluctuate according to complex deterministic
patterns \cite{mori1}. Under these conditions it is of interest to check up
whether the statistical-mechanical structure subsists, and if so, examine if
it is unchanged or if it has acquired a new form.

With this purpose in mind, the exploration of possible limits of validity of
the canonical statistical mechanics, an ideal model system is a
one-dimensional map at the transition between chaotic and regular behavior,
represented by well-known critical attractors, such as the Feigenbaum
attractor. So far, recent studies \cite{robledo1} have concentrated on the
dynamics \textit{inside} the attractor and have revealed that these
trajectories obey remarkably rich scaling properties not known previously at
this level of detail \cite{robledo2}. The results are exact and clarify \cite%
{robledo2} the relationship between the original modification \cite{politi1,mori1} 
of the thermodynamic approach to chaotic attractors \cite{thermo1,thermo2,thermo3} 
for this type of incipiently chaotic
attractor, and some aspects of the $q$-deformed statistical mechanical
formalism \cite{note1,robledo3,tsallis1}. The complementary
part of the dynamics, that of advance \textit{on the way} to the attractor,
has, to our knowledge, not been analyzed, nor understood, with a similar
degree of thoroughness. The process of convergence of trajectories into the
Feigenbaum attractor poses several interesting questions that we attempt to
answer here and elsewhere \cite{robledo00} based on the comprehensive
knowledge presented below. Prominent amongst these questions is the nature
of the connection between the two sets of dynamical properties, within and
outside the attractor. As it turns out, these two sets of properties are
related to each other in a statistical-mechanical manner, i.e. the dynamics 
\textit{at} the attractor provides the configurations in a partition
function while the \textit{approach} to the attractor is described by an
entropy obtained from it. As we show below, this statistical-mechanical
property conforms to a $q$-deformation \cite{note1} of the ordinary
exponential weight statistics.

Trajectories inside the attractor visit positions forming oscillating
log-periodic patterns of ever increasing amplitude. However, when the
trajectories are observed only at specified times, positions align according
to power laws, or $q$-exponential functions that share the same $q$-index
value \cite{robledo2,robledo3}. Further, all such sequences of
positions can be shifted and seen to collapse into a single one by a
rescaling operation similar to that observed for correlations in glassy
dynamics, a property known as ``aging'' \cite{robledo3,robledo4}. The
structure found in the dynamics is also seen to consist of a family of
Mori's $q$-phase transitions \cite{mori1}, via which the connection is made
between the modified thermodynamic approach and the $q$-statistical property
of the sensitivity to initial conditions \cite{robledo2,robledo3}. On
the other hand, a foretaste of the nature of the dynamics outside the
critical attractor can be appreciated by considering the dynamics towards
the so-called superstable cycles, or supercycles, the family of periodic attractors with Lyapunov
exponents that diverge towards minus infinity. This infinite family of
attractors has as accumulation point the transition to chaos, which for the
period-doubling route is the Feigenbaum attractor. As described below, the
basins of attraction for the different positions of the cycles develop
fractal boundaries of increasing complexity as the period-doubling structure
advances towards the transition to chaos. The fractal boundaries, formed by
the preimages of the repellor, display hierarchical structures organized
according to exponential clusterings that manifest in the dynamics as
sensitivity to the final state and transient chaos. The hierarchical
arrangement expands as the period of the supercycle increases.

To observe the general procedure followed by trajectories to reach the
attractors, and their complementary repellors, we consider an ensemble of
uniformly distributed initial conditions $x_{0}$ spanning the entire phase
space interval. We find that this is a highly structured process encoded in
sequences of positions shared by as many trajectories with different $x_{0}$%
. Clearly, there is always a natural dynamical ordering in the $x_{0}$ as
any trajectory of length $t$ contains consecutive positions of other
trajectories of lengths $t-1$, $t-2$, etc. with initial conditions $%
x_{0}^{\prime }$, $x_{0}^{\prime \prime }$, etc. that are images under
repeated map iterations of $x_{0}$. In the case of the Feigenbaum attractor
the initial conditions form two sets, dense in each other, of preimages of
each the attractor and the repellor. There is an infinite-level structure
within these sets that, as we shall see, is reflected by the infinite number
of families of phase-space gaps that complement the multifractal layout of
both attractor and repellor. These families of gaps appear sequentially in
the dynamics, beginning with the largest and followed by other sets
consisting of continually increasing elements with decreasing widths. The
number of gaps in each set of comparable widths increases as $2^{k}$, $%
k=0,1,\ldots $ and their widths can be ordered according to power laws of
the form $\alpha ^{-k}$, where $\alpha $ is Feigenbaum's universal constant $%
\alpha \simeq 2.5091$. We call $k$ the order of the gap set. Furthermore, by
considering a fine partition of phase-space, we determine the overall rate
of approach of trajectories towards the attractor (and to the repellor).
This rate is measured by the fraction of bins $W(t)$ still occupied by
trajectories at time $t$ \cite{lyra1}. The power law with log-periodic
modulation displayed by $W(t)$ \cite{lyra1} is explained in terms of the
progression of gap formation, and its self-similar features are seen to
originate in the unlimited hierarchy feature of the preimage structure.

Before proceeding to give details in the following sections of the
aforementioned dynamics, we recall basic features of the bifurcation forks
that form the period-doubling cascade sequence in unimodal maps, epitomized
by the logistic map $f_{\mu }(x)=1-\mu x^{2}$, $-1\leq x\leq 1$, $0\leq \mu
\leq 2$ \cite{schuster1,beck1}. The superstable periodic orbits of
lengths $2^{N}$, $N=1,2,3,\ldots $, are located along the bifurcation forks,
i.e. the control parameter value $\mu =\overline{\mu }_{N}<\mu _{\infty }$
for the superstable $2^{N}$-attractor is that for which the orbit of period $%
2^{N}$ contains the point $x=0$, where $\mu _{\infty }=1.401155189\ldots $
is the value of $\mu $ at the period-doubling accumulation point. The
positions (or phases) of the $2^{N}$-attractor are given by $x_{m}=f_{%
\overline{\mu }_{N}}^{(m)}(0)$, $m=1,2,\ldots ,2^{N}$. Notice that
infinitely many other sequences of superstable attractors appear at the
period-doubling cascades within the windows of periodic attractors for
values of $\mu >$ $\mu _{\infty }$. Associated with the $2^{N}$-attractor at $%
\mu =\overline{\mu }_{N}$ there is a $(2^{N}-1)$-repellor consisting of $%
2^{N}-1$ positions $y_{m}$, $m=1,2,\ldots ,2^{N}-1$. These positions are the
unstable solutions $\left\vert df_{\overline{\mu }_{N}}^{(2^{k})}(y)/dy%
\right\vert <1$ of $y=f_{\overline{\mu }_{N}}^{(2^{k})}(y)$, $%
k=0,1,2,\ldots ,N-1$. The first, $k=0$, originates at the initial
period-doubling bifurcation, the next two, $k=1$, start at the second
bifurcation, and so on, with the last group of $2^{N-1}$, $k=N-1$, stemming
from the $N-1$ bifurcation. We find it useful to order the repellor
positions, or simply, repellors, present at $\mu =\overline{\mu }_{N}$,
according to a hierarchy or tree, the ``oldest'' with $k=0$ up to the most
``recent'' ones with $k=N-1$. The repellors' order is given by the value of $k$.
Finally, we define the preimage $x^{(j)}$ of order $j$ of position $x$ to
satisfy $x=h^{(j)}(x^{(j)})$ where $h^{(j)}(x)$ is the $j$-th composition of
the map $h(x)\equiv f_{\overline{\mu }_{N}}^{(2^{N-1})}(x)$. We have omitted
reference to the $2^{N}$-cycle in $x^{(j)}$ to simplify the notation. The
interval lengths or diameters $d_{N,m}$ are measured when considering the
superstable periodic orbits of lengths $2^{N}$. The $d_{N,m}$ are defined
(here) as the (positive) distances of the elements $x_{m}$, $m=0,1,2,\ldots
,2^{N}-1$, to their nearest neighbors $f_{\overline{\mu }%
_{N}}^{(2^{N-1})}(x_{m})$, i.e. 
\begin{equation}
d_{N,m}\equiv \left\vert f_{\overline{\mu }_{N}}^{(m+2^{N-1})}(0)-f_{%
\overline{\mu }_{N}}^{(m)}(0)\right\vert .
\end{equation}%
For large $N$, $d_{N,0}/d_{N+1,0}\simeq \alpha $. We present explicit
results for the logistic map, which has a quadratic maximum, but the results
are easily extended to unimodal maps with general nonlinearity $z>1$.

\begin{figure}[h!]
\centering
\includegraphics[width=.5\textwidth]{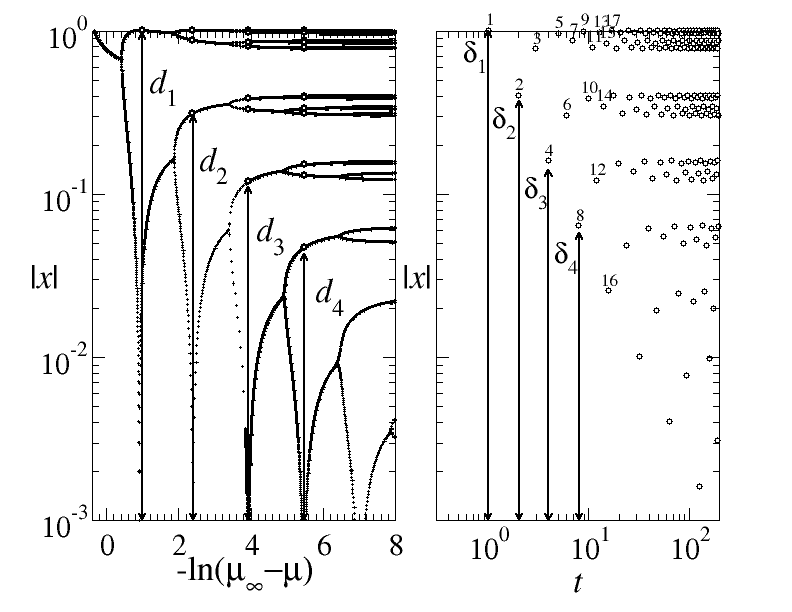}
\caption{ 
{\small \ Left panel: Absolute value of attractor positions for the
logistic map }${\small f}_{\mu }{\small (x)}${\small \ in logarithmic scale
as a function of the logarithm of the control parameter difference }${\small %
\mu }_{\infty }{\small -\mu }${\small . Right panel: Absolute value of
trajectory positions for the logistic map }${\small f}_{\mu }{\small (x)}$%
{\small \ at }${\small \mu }_{\infty }${\small \ with initial condition }$%
{\small x}_{0}{\small =0}${\small \ in logarithmic scale as a function of
the logarithm of time }$t${\small , also shown by the numbers close to the
circles. The arrows indicate the equivalence between the diameters }$\delta_{N}$%
{\small \ in the left panel, and position differences }$\delta_{N}${\small \ with
respect to }${\small x}_{0}{\small =0}${\small \ in the right panel.}
}
\label{fig.1.1}
\end{figure}

Central to our discussion is the following broad property: Time evolution at 
$\mu _{\infty }$ from $t=0$ up to $t\rightarrow \infty $ traces the
period-doubling cascade progression from $\mu =0$ up to $\mu _{\infty }$.
Not only is there a close resemblance between the two developments but also
quantitative agreement. For instance, the trajectory inside the Feigenbaum
attractor with initial condition $x_{0}=0$, the $2^{\infty }$-supercycle
orbit, takes positions $x_{t}$ such that the distances between appropriate
pairs of them reproduce the diameters $d_{N,m}$ defined from the supercycle
orbits with $\overline{\mu }_{N}<\mu _{\infty }$. See Fig. \ref{fig.1.1}, where the
absolute value of positions and logarithmic scales are used to illustrate
the equivalence. This property has been key to obtain rigorous results for
the sensitivity to initial conditions for the Feigenbaum attractor \cite{robledo1,robledo3}.

The layout of the rest of this paper is the following. In Sec. \ref{sec2} we
present the dynamical properties of the family of supercycles, describing
their preimage structure, final state sensitivity and transient chaos.
Details of the superstrong insensitivity to initial conditions displayed by
these attractors are given as an Appendix. In Sec. \ref{sec3} we make use of the
results of the previous section to describe the dynamical properties of
approach to the Feigenbaum attractor. We provide details of its preimage
structure, the sequential opening of phase-space gaps, and the scaling for
the trajectories' rate of convergence to the attractor (and repellor). In
Sec. \ref{sec4} we explain the aforesaid statistical-mechanical structure
lying beneath the dynamics of an ensemble of trajectories en route to the
Feigenbaum attractor (and repellor). In Sec. \ref{sec5} we summarize our
results.


\section{Hierarchical properties in the dynamics of supercycle attractors}
\label{sec2}
To obtain dynamical properties with previously unstated detail we determined
the organization of the \textit{entire} set of trajectories as generated by
all possible initial conditions. We find that the paths taken by the full
set of trajectories in their way to the supercycle attractors (or to their
complementary repellors) are far from unstructured. The preimages of the
attractor of period $2^{N}$, $N=1,2,3,\ldots $ are distributed into
different basins of attraction, one for each of the $2^{N}$\ phases
(positions) that compose the\ cycle. When $N\geq 2$ these basins are
separated by fractal boundaries whose complexity increases with increasing $%
N $. The boundaries consist of the preimages of the corresponding repellor
and their positions cluster around the $2^{N}-1$ repellor positions
according to an exponential law. As $N$ increases the structure of the basin
boundaries becomes more involved. That is, the boundaries for the $2^{N}$
cycle develop new features around those of the previous $2^{N-1}$ cycle
boundaries, with the outcome that a hierarchical structure arises, leading
to embedded clusters of clusters of boundary positions, and so forth.

The dynamics associated with families of trajectories always displays a
distinctively concerted order that reflects the repellor preimage boundary
structure of the basins of attraction. That is, each trajectory has an
initial condition that is identified as an attractor (or repellor) preimage
of a given order, and this trajectory necessarily follows the steps of other
trajectories with initial conditions of lower preimage order belonging to a
given chain or pathway to the attractor (or repellor). This feature gives
rise to transient chaotic behavior different from that observed at the last
stage of approach to the attractor. When the period $2^{N}$ of the cycle
increases the dynamics becomes more involved with increasingly more complex
stages that reflect the preimage hierarchical structure. As a final point,
shown in the appendix, in the closing part of the last leg of the
trajectories an ultra-rapid convergence to the attractor is observed, with a
sensitivity to initial conditions that decreases as an exponential of an
exponential in time. In relation to this we find that there is a functional
composition renormalization group (RG) fixed-point map associated with the
supercycle attractor, and this can be expressed in closed form by the same
kind of $q$ -exponential function found for both the pitchfork and tangent
bifurcation attractors \cite{robledo5,robledo6}, like that
originally derived by Hu and Rudnick for the latter case \cite{hu1}.

\subsection{Preimage structure of supercycle attractors}

\begin{figure}[h!]
\centering
\includegraphics[width=.5\textwidth]{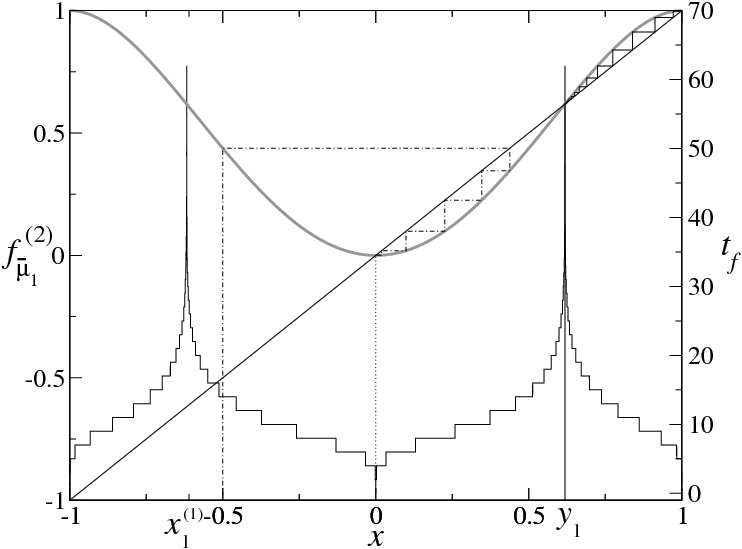}
\caption{ {\protect\small Left axis: The twice iterated map }$%
{\protect\small f}_{\overline{\protect\mu }_{1}}^{(2)}{\protect\small (x)}$, with 
$\overline{\protect\mu }_{1}{\protect\small =1}${\protect\small \ (gray
line)}.{\protect\small \ Right axis: Time of flight }${\protect\small t}_{f}%
{\protect\small (x)} ${\protect\small , the number of iterations necessary
for a trajectory with initial condition at }${\protect\small x}$%
{\protect\small \ to reach an attractor position. The values of }$%
{\protect\small x}${\protect\small \ near the high spikes correspond to
initial conditions very close to the repellor and its preimage. We present
three example trajectories (and the }${\protect\small y=x}${\protect\small \
line as an aid to visualize them): The dotted line shows a trajectory that
starts at the attractor position }${\protect\small x=0}${\protect\small \
and remains there. The solid line is a trajectory starting near the repellor
at }${\protect\small y}_{{\protect\small 1}}${\protect\small , and after a
large number of iterations reaches the attractor position }${\protect\small %
x=1}${\protect\small . Finally, the dash-dotted line is an orbit starting at 
}${\protect\small x=0.5}${\protect\small \ that in just a few iterations
reaches }${\protect\small x=0}${\protect\small .} }
\label{fig.1.2}
\end{figure}

The core source of our description of the dynamics towards the supercycle
attractors is a measure of the relative ``time of flight'' $t_{f}(x_{0})$ for
a trajectory with initial condition $x_{0}$ to reach the attractor. The
function $t_{f}(x_{0})$ is obtained for an ensemble representative of 
\textit{all} initial conditions $-1\leq x_{0}\leq 1$. This comprehensive
information is determined through the numerical realization of every
trajectory, up to a small cutoff $\varepsilon >0$ at its final stage. The
cutoff $\varepsilon $ considers a position $x_{f}\leq x_{m}\pm \varepsilon $
to be effectively the attractor phase $x_{m}$. This, of course, introduces
an approximation to the real time of flight, which can be arbitrarily large
for those $x_{0}$ close to a repellor position $y_{m}$ or close to any of
its infinitely many preimages, $x_{m}^{(j)}$, $j=1,2,\ldots $, $N\geq 2$. In
such cases the finite time $t_{f}(x_{0};\varepsilon )$ can be seen to
diverge $t_{f}\rightarrow \infty $ as $x_{0}\rightarrow y_{m}$ and $%
\varepsilon \rightarrow 0$. As a simple illustration, in Fig. \ref{fig.1.2} we
show the time of flight $t_{f}(x_{0})$ for the period $2$ supercycle at $\mu
=\overline{\mu }_{1}$ with $\varepsilon =10^{-9}$ together with the
twice-composed map $f_{\overline{\mu }_{1}}^{(2)}(x)$ and a few
representative trajectories. We observe two peaks in $t_{f}(x_{0};%
\varepsilon )$ at $y_{1}=(-1+\sqrt{1+4\overline{\mu }_{1}})/2\overline{\mu }%
_{1}\simeq 0.6180340\ldots $, the fixed-point repellor and at its (only)
preimage $x_{1}^{(1)}=-y_{1}$, where $y_{1}=f_{\overline{\mu }%
_{1}}^{(2)}(x_{1}^{(1)})$. Clearly, there are two basins of attraction each
for the two positions or phases, $x_{1}=0$ and $x_{2}=1$, of the attractor.
For the former it is the interval $x_{1}^{(1)}<x_{0}<y_{1}$, whereas for the
second one it consists of the intervals $-1\leq x_{0}<x_{1}^{(1)}$ and $%
y_{1}<x_{0}\leq 1$. The multiple-step structure of $t_{f}(x_{0})$, of four
time units each step, reflects the occurrence of intervals of initial
conditions with common attractor phase preimage order $k$.

\begin{figure}[h!]
\centering
\includegraphics[width=.5\textwidth]{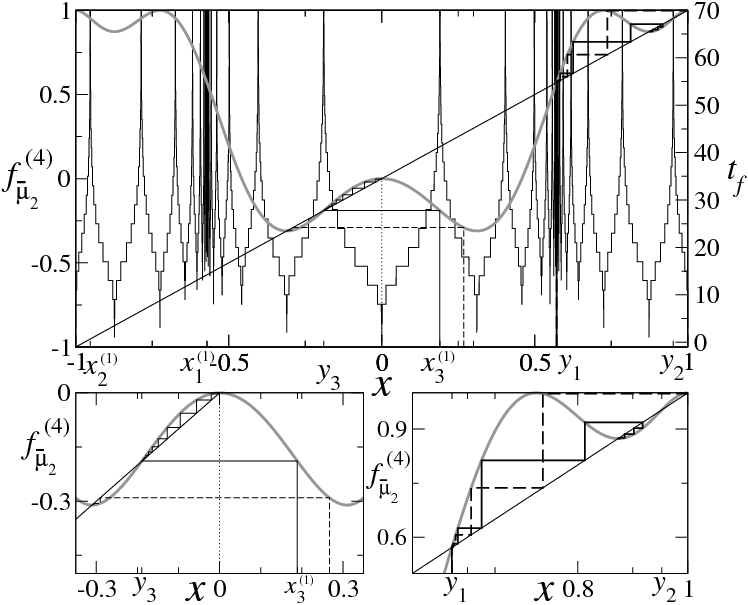}
\caption{ {\protect\small Same as Fig. \protect\ref{fig.1.2} but for }$f_{%
\overline{\protect\mu }_{2}}^{(4)}(x)$, $\overline{\protect\mu }_{2}\simeq 
{\protect\small 1.31070264}${\protect\small . Two orbits (solid bold and
dashed bold lines) start very near each other and by the position }$%
{\protect\small y}_{{\protect\small 1}}${\protect\small \ of the old
repellor, although indistinguishable at the beginning they take very different
subsequent paths to reach the attractor positions at }${\protect\small x=1}$%
{\protect\small \ and }${\protect\small x\simeq 0.8734} ${\protect\small .
See bottom right panel. Another two orbits (in dotted and solid lines) start
at an attractor position }${\protect\small x=1}${\protect\small \ and at a
repellor preimage position, respectively. Finally, one more orbit (dashed line) 
starts at an intermediate initial condition and reaches very
quickly the attractor position at }${\protect\small x\simeq -0.310703}$%
{\protect\small . See bottom left panel.} }
\label{fig.1.3}
\end{figure}

\begin{figure}[h!]
\centering
\includegraphics[width=.5\textwidth]{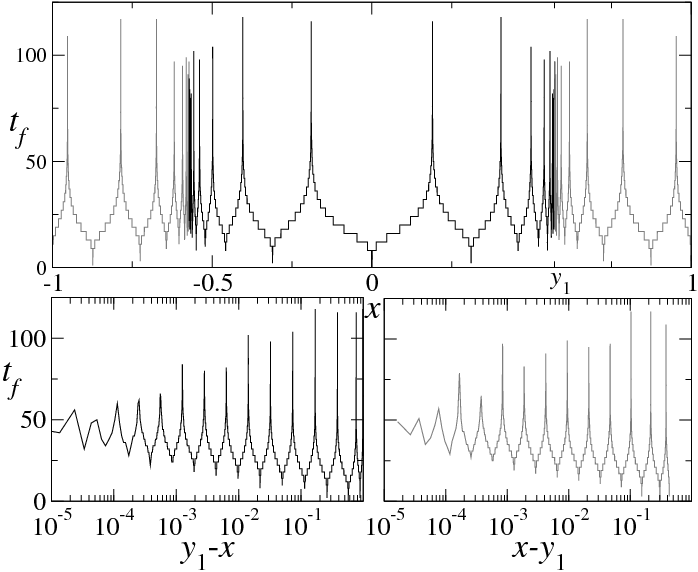}
\caption{ {\protect\small Top panel: Time of flight }${\protect\small t}_{f}%
{\protect\small (x)}${\protect\small \ for }${\protect\small N=2}$%
{\protect\small \ (as in Fig. 3), the black lines correspond to initial
conditions that terminate at the attractor positions }${\protect\small x=0}$%
{\protect\small \ and }${\protect\small x\simeq -0.310703}${\protect\small ,
and the gray lines to trajectories ending at }${\protect\small x=1}$%
{\protect\small \ and }${\protect\small x\simeq 0.8734}${\protect\small .
Right (left) bottom panel: Same as top panel, but plotted against the
logarithm of }${\protect\small x-y}_{{\protect\small 1}}${\protect\small \ (}%
${\protect\small y}_{{\protect\small 1}}{\protect\small -x}${\protect\small %
). It is evident that the peaks are arranged exponentially around the old
repellor position }${\protect\small y}_{{\protect\small 1}}${\protect\small %
, i.e., they appear equidistant on a logarithmic scale. See Fig. \protect\ref%
{fig.1.5}.} }
\label{fig.1.4}
\end{figure}

For the next supercycle---period $4$---the preimage structure turns out to
be a good deal more involved than the straightforward structure for $%
\overline{\mu }_{1}$. In Fig. \ref{fig.1.3} we show the times of flight $%
t_{f}(x_{0})$ for the $N=2$ supercycle at $\mu =\overline{\mu }_{2}$ with $%
\varepsilon =10^{-9}$, the map $f_{\overline{\mu }_{2}}^{(4)}(x)$ is
superposed as a reference to indicate the four phases of the attractor (at $%
x_{1}=0$, $x_{2}=1$, $x_{3}\simeq -0.3107\ldots$, and $x_{4}\simeq
0.8734\ldots$) and the three repellor positions (at $y_{1}\simeq
0.5716635\ldots$, $y_{2}\simeq 0.952771\ldots$, and $y_{3}\simeq
-0.189822\ldots$) In Fig. \ref{fig.1.3} there are also shown four trajectories
each of which terminates at a different attractor phase. We observe a
proliferation of peaks and valleys in $t_{f}(x_{0})$, actually, an infinite
number of them, that cluster around the repellor at $y_{1}\simeq
0.5716635\ldots$ and also at its preimage at $x_{1}^{(1)}=-y_{1}$ (these are
the positions at $\mu = \overline{\mu }_{2}$ of the ``old'' repellor and its
preimage in the previous $N=1$ case). Notice that the steps in the valleys
of $t_{f}(x_{0})$ are now eight time units each. The nature of the
clustering of peaks (repellor phase preimages) and the bases of the valleys
(attractor phase preimages) is revealed in Fig. \ref{fig.1.4} where we plot $%
t_{f}$ on a logarithmic scale for the variables $\pm (x-y_{1})$. There is an
exponential clustering of the preimage structure around both the old
repellor and its preimage. This scaling property is corroborated in Fig. \ref%
{fig.1.5} from which we obtain $x-y_{1}\simeq 7.5\times10^{-5}\exp (0.80\,l)$,
where $l=1,2,3, \ldots$ is a label for consecutive repellor preimages.

\begin{figure}[h!]
\centering
\includegraphics[width=.5\textwidth]{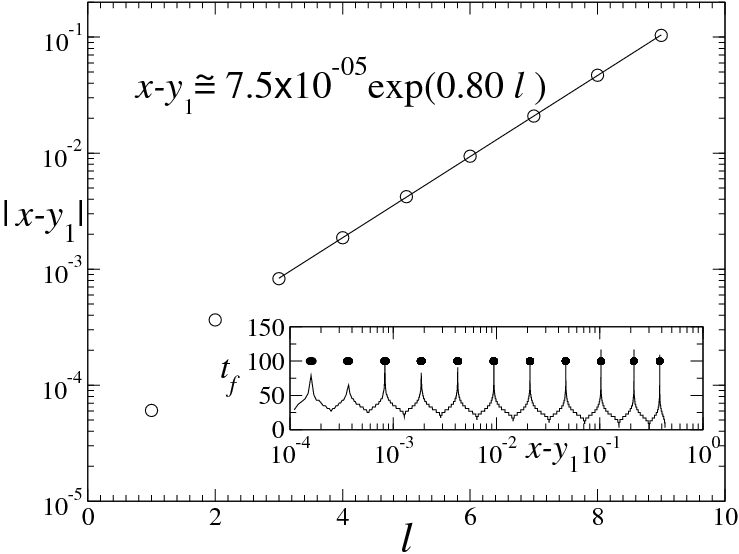}
\caption{ {\protect\small Corroboration of preimage exponential clustering
around the repellor position }${\protect\small y}_{{\protect\small 1}}\simeq 
{\protect\small 0.571663}${\protect\small \ when }${\protect\small N=2}$%
{\protect\small \ and }$\overline{\protect\mu }_{2}\simeq {\protect\small %
1.31070264}$.{\protect\small \ The variable }${\protect\small l}$%
{\protect\small \ labels consecutive equidistant peaks in the inset. The
peaks correspond to preimages of the repellor. See text.} }
\label{fig.1.5}
\end{figure}

\begin{figure}[h!]
\centering
\includegraphics[width=.5\textwidth]{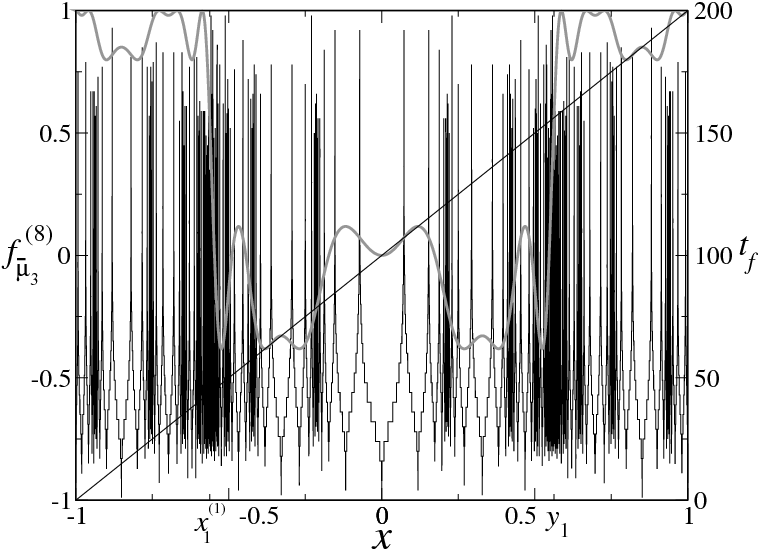}
\caption{ {\protect\small Same as Fig. \protect\ref{fig.1.3} but for }$%
{\protect\small f}_{\overline{\protect\mu }_{3}}^{(8)}{\protect\small (x)}$, 
$\overline{\protect\mu }_{3}{\protect\small \simeq 1.38154748}$.%
{\protect\small \ Here }${\protect\small y}_{{\protect\small 1}}%
{\protect\small =0.56264475}${\protect\small .} }
\label{fig.1.6}
\end{figure}

\begin{figure}[h!]
\centering
\includegraphics[width=.5\textwidth]{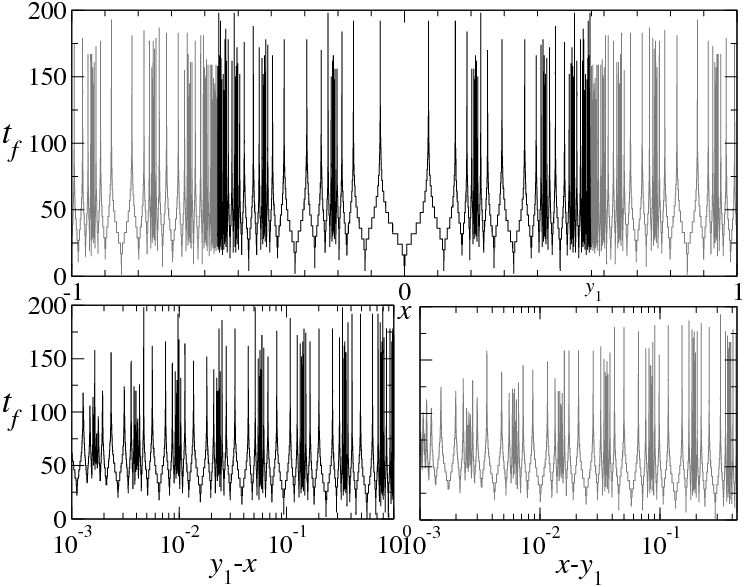}
\caption{ {\protect\small Same as Fig. \protect\ref{fig.1.4} but for }$%
{\protect\small N=3}${\protect\small . The black lines correspond to initial
conditions that terminate at any of the four attractor positions close or
equal to }${\protect\small x=0}${\protect\small , and the gray lines to
trajectories ending at any of the other four attractor positions close or
equal to }${\protect\small x=1}$. {\protect\small As the bottom panels show,
on a logarithmic scale, in this case there are (infinitely) many clusters of
peaks (repellor preimages) equidistant from each other.} }
\label{fig.1.7}
\end{figure}

\begin{figure}[h!]
\centering
\includegraphics[width=.5\textwidth]{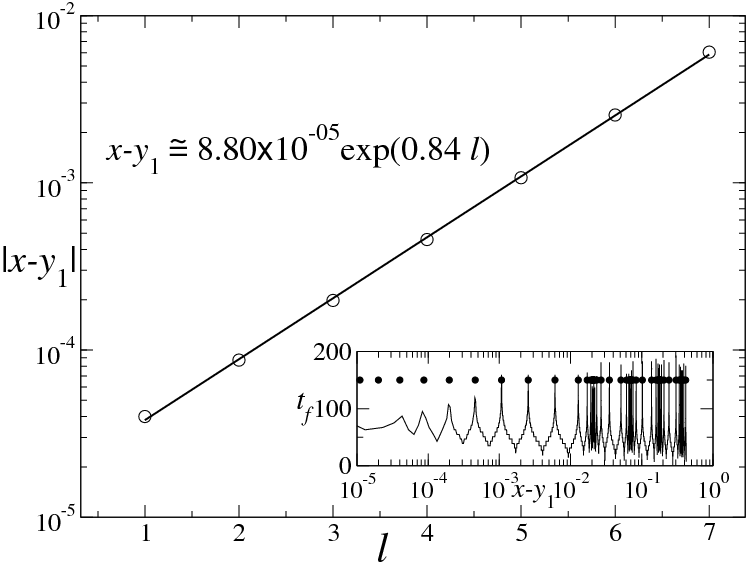}
\caption{ {\protect\small Same as Fig. \protect\ref{fig.1.5} but for }$%
{\protect\small N=3}${\protect\small .} }
\label{fig.1.8}
\end{figure}

A comparable leap in the complexity of the preimage structure is observed
for the following---period $8$---supercycle. In Fig. \ref{fig.1.6} we show $%
t_{f}(x_{0})$ for the $N=3$ supercycle at $\mu =\overline{\mu }_{3}$ with $%
\varepsilon =10^{-9}$, together with the map $f_{\overline{\mu }%
_{2}}^{(8)}(x)$ placed as reference to facilitate the identification of the
locations of the eight phases of the attractor, $x_{1}$ to $x_{8}$, and the
seven repellor positions, $y_{1}$ to $y_{7}$. In addition to a huge proliferation
of peaks and valleys in $t_{f}(x_{0})$, we observe now the development of
clusters of clusters of peaks centered around the repellor at $y_{1}\simeq
0.5626447\ldots$ and its preimage at $x_{1}^{(1)}=-y_{1}$ (these are now the
positions of the original repellor and its preimage for the $N=1$ case when $%
\mu =\overline{\mu }_{3}$). The steps in the valleys of $t_{f}(x_{0})$ have
become 16 time units each. Similarly to the clustering of peaks
(repellor phase preimages) and valleys (attractor phase preimages) for the
previous supercycle at $\overline{\mu }_{2}$, the spacing arrangement of the
new clusters of clusters of peaks is determined in Fig. \ref{fig.1.7} where we
plot $t_{f}$ in a logarithmic scale for the variables $\pm (x-y_{1})$. In
parallel to the previous cycle an exponential clustering of clusters of the
preimage structure is found around both the old repellor and its preimage.
This scaling property is quantified in Fig. \ref{fig.1.8} from which we obtain $%
x-y_{1}\simeq 8.8\times 10^{-5}\exp (0.84\,l),$ where $l=1,2,3, \ldots$
counts consecutive clusters.

\begin{figure}[h!]
\centering
\includegraphics[width=.5\textwidth]{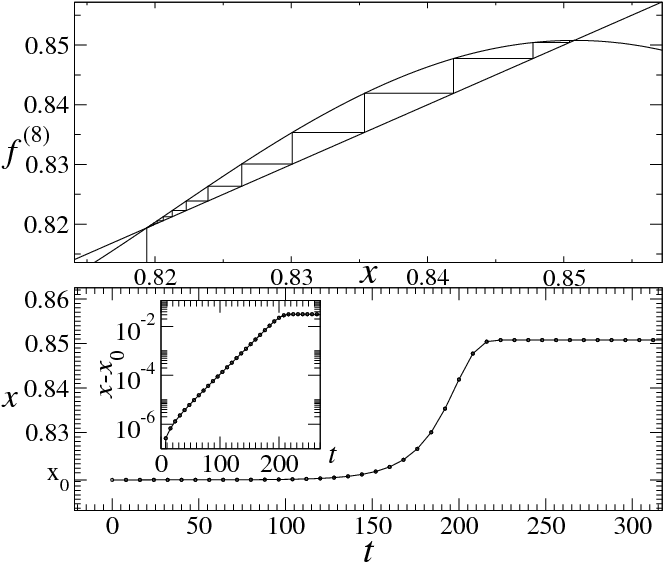}
\caption{ {\protect\small Top panel: Detail of the map }${\protect\small f}_{%
\overline{\protect\mu }_{3}}^{(8)}{\protect\small (x)}${\protect\small \ and
a trajectory in the proximity of the attractor position located at }$%
{\protect\small x\simeq 0.850780}$. {\protect\small The trajectory
originated very close to the repellor position at }${\protect\small %
x=0.819378}${\protect\small . Bottom panel: The same trajectory of the
eight-iterated map as a function of time }${\protect\small t}$%
{\protect\small . Inset: Corroboration of the exponential nature of the
trajectory after it leaves the repellor and before the final approach to the
attractor position.} }
\label{fig.1.9}
\end{figure}

An investigation of the preimage structure for the next $N=4$ supercycle at $%
\mu =\overline{\mu }_{4}$ leads to another substantial increment in the
complications of the structure of the preimages but with such density that is 
cumbersome to describe here. Nevertheless it is clear that the main
characteristic in the dynamics is the development of a hierarchical
organization of the preimage structure as the period $2^{N}$ of the
supercycles increases.

\subsection{Final state sensitivity and transient chaos}

With the knowledge gained about the features displayed by the times of
flight $t_{f}(x_{0})$ for the first few supercycles it is possible to
determine and understand how the leading properties in the dynamics of
approach to these attractors arise. The information contained in $%
t_{f}(x_{0})$ can be used to demonstrate in detail how the concepts of final
state sensitivity \cite{grebogi1}---due to attractor multiplicity---and
transient chaos \cite{beck1}---prevalent in the presence of repellors that
coexist with periodic attractors---are realized in a given dynamics. Final state
sensitivity is the consequence of fractal boundaries separating coexisting
attractors. In our case there is always a single attractor but its positions
or phases play an equivalent role \cite{grebogi2}. Transient chaos \cite%
{beck1} is due to fast separation in time of nearby trajectories by the
action of a repellor and results in a sensitivity to initial conditions that
grows exponentially up to a crossover time after which decay sets in. We
describe how both properties result from an extremely ordered flow of
trajectories towards the attractor. This order is imprinted by the preimage
structure described above.

For the simplest supercycle at $\mu =\overline{\mu }_{1}$ there is trivial
final state sensitivity as the boundary between the two basins of the
phases, $x_{1}=0$ and $x_{2}=1$, consists only of the two positions $%
y_{1}\simeq 0.5716635\ldots$ and $x_{1}^{(1)}=-y_{1}$. See Fig. \ref{fig.1.2}.
Consider the length $\delta $ of a small interval around a given value of $%
x_{0}$ containing either $y_{1}$ or $x_{1}^{(1)}$, when $\delta \rightarrow
0 $ any uncertainty as to the final phase of the trajectory disappears. It
is also simple to verify that when $x_{0}$ is close to $y_{1}$ or $%
x_{1}^{(1)}$ the resulting trajectories increase their separation at initial
and intermediate times displaying transient chaos in a straightforward
fashion. In Fig. \ref{fig.1.9} we show the same type of transitory exponential
sensitivity to initial conditions for a trajectory at $\mu =\overline{\mu }%
_{3}$ after it reaches a repellor position in the final journey towards the
period $8$ attractor. This behavior is common to all periodic attractors
when trajectories come near a repellor at the final leg of their journey.

\begin{figure}[h!]
\centering
\includegraphics[width=.5\textwidth]{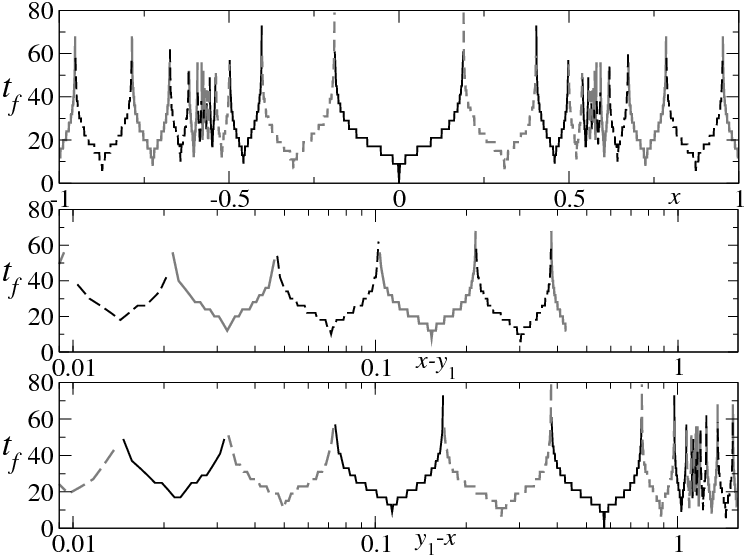}
\caption{ {\protect\small Time of flight }${\protect\small t}_{f}%
{\protect\small (x)}${\protect\small \ for }${\protect\small N=2}$ 
{\protect\small (equivalent to Fig. \protect\ref{fig.1.4}). Different types of
lines are used for different sub-basins of attraction that form the fractal
boundary between the attractor positions. The black solid line corresponds
to the attractor position }${\protect\small x=0}${\protect\small , the gray
solid line to }${\protect\small x=1}${\protect\small , the dashed gray line to }$%
{\protect\small x\simeq -0.310703}${\protect\small , and the dashed black line
to }${\protect\small x\simeq 0.873470}${\protect\small . The sub-basins are
separated in an alternating fashion by repellor preimages (peaks) and there
are just two types of sub-basins depending on }${\protect\small |x|}$%
{\protect\small \ being larger or smaller than }${\protect\small y}_{%
{\protect\small 1}}${\protect\small .} }
\label{fig.1.10}
\end{figure}

For $\mu =\overline{\mu }_{2}$ there are more remarkable properties arising
from the more complex preimage structure. There is a concerted migration of
initial conditions seeping through the boundaries between the four basins of
attraction of the phases. These boundaries, shown in Fig. \ref{fig.1.10}, form
a fractal network of interlaced sub-basins separated from each other by two
preimages of different repellor phases and have at their bottom a preimage
of an attractor phase. Trajectories on one of these sub-basins move to the
nearest sub-basin of its type (next-nearest neighbor in actual distance in
Fig. \ref{fig.1.10}) at each iteration of the map $f_{\overline{\mu }%
_{2}}^{(4)} $ (four time steps for the original map $f_{\overline{\mu }_{2}}$%
). The movement is always away from the center of the cluster at the old
repellor position $y_{1}$ or at its preimage $x_{1}^{(1)}$ (located at the
steepest slope inflection points of $f_{\overline{\mu }_{2}}^{(4)}$ shown in
Fig. \ref{fig.1.3}). Once a trajectory is out of the cluster (contained between
the maxima and minima of $f_{\overline{\mu }_{2}}^{(4)}$ next to the
mentioned inflection points) it proceeds to the basin of attraction of an
attractor phase (separated from the cluster by the inflection points with
gentler slope of $f_{\overline{\mu }_{2}}^{(4)}$ in Fig. \ref{fig.1.3}) where
its final stage takes place. When we consider a large ensemble of initial
positions, distributed uniformly along all phase-space, the common journey
towards the attractor displays an exceedingly ordered pattern. Each initial
position $x_{0}$ within either of the two clusters of sub-basins is a
preimage of a given order $k$ of a position in the main basin of attraction.
Each iteration of $f_{\overline{\mu }_{2}}^{(4)}$ reduces the order of the
preimage from $k$ to $k-1$, and the new position $x_{0}^{\prime }=f_{%
\overline{\mu }_{2}}^{(4)}(x_{0})$ replaces the initial position $%
x_{0}^{\prime }$ (a preimage of order $k-1$) of another trajectory that
under the same time step has migrated to the initial position $x_{0}^{\prime
\prime }$ (a preimage of order $k-2$) of another trajectory, and so on.

\begin{figure}[h!]
\centering
\includegraphics[width=.5\textwidth]{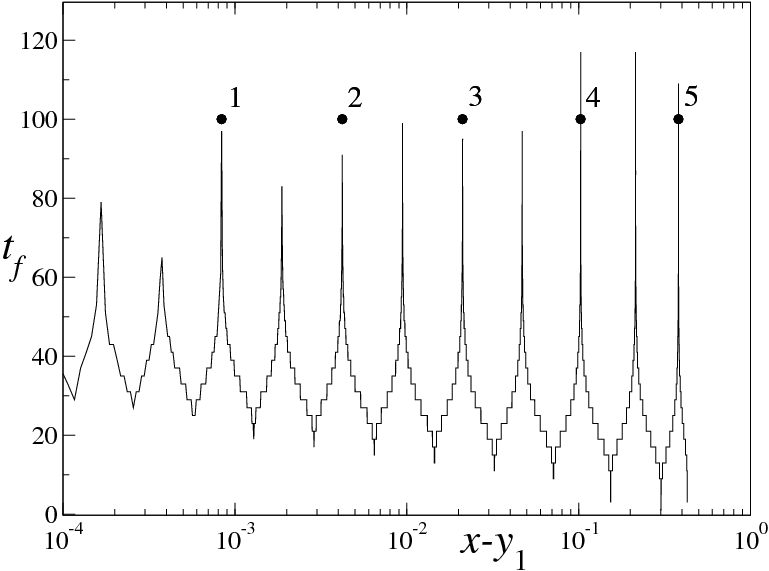}
\caption{ {\protect\small Evolution of an orbit starting very close to a
peak (repellor preimage) inside a cluster when }${\protect\small N=2}$%
{\protect\small , }$\overline{\protect\mu }_{2}\simeq {\protect\small %
1.31070264}${\protect\small \ and }${\protect\small y}_{1}{\protect\small %
=0.571663}${\protect\small . (See Fig. \protect\ref{fig.1.4}.) The trajectory
describes an exponential in time moving away from }${\protect\small y}_{%
{\protect\small 1}}${\protect\small . In the figure, positions of the
trajectory correspond to the black solid dots, and iterations correspond to
the number associated with each dot.} }
\label{fig.1.11}
\end{figure}

It is clear that the dynamics at $\mu =\overline{\mu }_{2}$ displays
sensitivity to the final state when the initial condition $x_{0}$ is located
near the core of any of the two clusters of sub-basins at $y_{1}$ and $%
x_{1}^{(1)}$ that form the boundary of the attractor phases. Any uncertainty
on the location of $x_{0}$ when arbitrarily close to these positions implies
uncertainty about the final phase of a trajectory. There is also transient
chaotic behavior associated to the migration of trajectories out of the
cluster as a result of the organized preimage resettlement mentioned above.
Indeed, the exponential disposition of repellor preimages shown in Fig. \ref%
{fig.1.5} is actually a realization of two trajectories with initial conditions
in consecutive peaks of the cluster structure. Therefore, the exponential
expression given in the previous section in relation to Fig. \ref{fig.1.5}
can be rewritten as the expression for a trajectory $x_{\theta }\simeq
x_{0}\exp (\lambda _{eff}\theta )$, with $x_{\theta }=x-y_{1}$, $x_{0}=$ $%
7.5\times 10^{-5}$, and $\lambda _{eff}=6.4$, where $\theta =1,2,3,\ldots $
Straightforward differentiation of $x_{\theta }$ with respect to $x_{0}$
yields an exponential sensitivity to initial conditions with positive
effective Lyapunov coefficient $\lambda _{eff}$. See Fig. \ref{fig.1.11}.

\begin{figure}[h!]
\centering
\includegraphics[width=.5\textwidth]{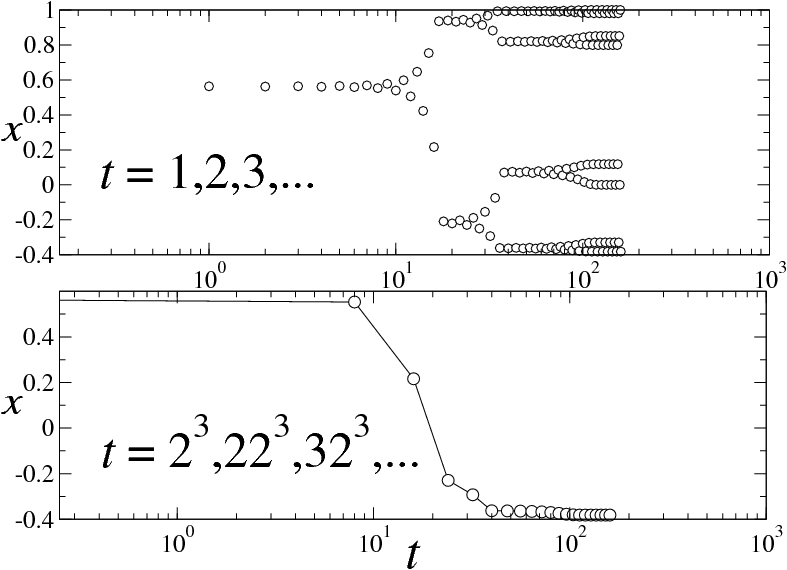}
\caption{ {\protect\small A trajectory for }$\overline{\protect\mu }_{3}%
{\protect\small \simeq 1.38154748}${\protect\small . Top panel: The circles
are positions for consecutive times in the iterations of the map }$%
{\protect\small f}_{\overline{\protect\mu }_{3}}{\protect\small (x)}$%
{\protect\small . The orbit starts very close to the old (period 1)
repellor }${\protect\small y}_{{\protect\small 1}}${\protect\small , then
moves close to a period 2 and subsequently to a period 4 repellor,
until finally it arrives at the period 8 attractor. Bottom
panel: Selection of a time subsequence of multiples of }${\protect\small 2}%
^{3}${\protect\small \ shows the same trajectory as an evolution towards one
particular attractor position or phase.} }
\label{fig.1.12}
\end{figure}

As can be anticipated, the dynamics of approach to the next supercycle at 
$\mu =\overline{\mu }_{3}$ can be explained by enlarging the description
presented above for $\mu =\overline{\mu }_{2}$ with the additional features
of its preimage structure already detailed in the preceding section. As
in the previous case, trajectories with initial conditions $x_{0}$ located
inside a cluster of sub-basins of the attractor phases will proceed to move
out of it in the systematic manner described for the only two isolated
clusters present when $\mu =\overline{\mu }_{2}$. However, now there is an
infinite number of such clusters arranged into two bunches that group
exponentially around the old repellor position $y_{1}$ and around its
preimage $x_{1}^{(1)}$. See Figs. \ref{fig.1.6}--\ref{fig.1.8}. Once such
trajectories leave the cluster under consideration they enter into a
neighboring cluster, and so forth, so that the trajectories advance out of
these fractal boundaries through the prolonged process of migration out of
the cluster of clusters before they proceed to the basins of the attraction
of the eight phases of this cycle. In Fig. \ref{fig.1.12} we show one such
trajectory in consecutive times $t=1,2,3,\ldots $ for the original map and
also in multiples of time $t=2^{3},22^{3},32^{3},\ldots $. The logarithmic
scale of the figure makes evident the retardation of each stage in the
process. As when $\mu =\overline{\mu }_{2}$, it is clear that in the
approach to the $\mu =\overline{\mu }_{3}$ attractor there is sensitivity to
the final state and transitory chaotic sensitivity to initial conditions.
Again, the exponential expression, given in the previous section
associated with the preimage structure of clusters of clusters of sub-basins, 
shown in Fig. \ref{fig.1.8}, can be interpreted as the expression of a
trajectory of the form $x_{\theta }\simeq x_{0}\exp (\lambda _{eff}\theta )$. 
Differentiation of $x_{\theta }$ with respect to $x_{0}$ yields again an
exponential sensitivity to initial conditions with positive effective
Lyapunov coefficient $\lambda _{eff}$.

When the period $2^{N}$ of the cycles increases we observe that the main
characteristic in the dynamics is the development of a hierarchical
organization in the flow of an ensemble of trajectories out of an
increasingly more complex disposition of the preimages of the attractor
phases. 


\section{Dynamical properties of approach to the Feigenbaum attractor}
\label{sec3}
\begin{figure}[h!]
\centering
\includegraphics[width=.5\textwidth]{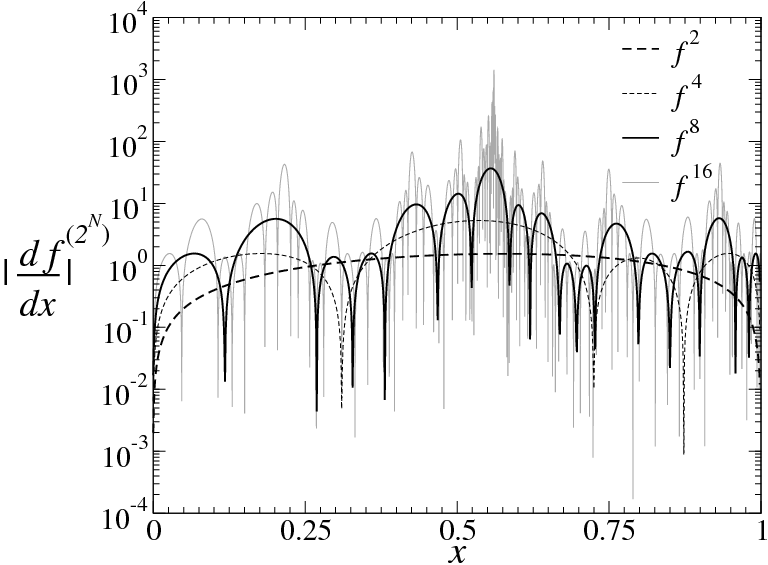}
\caption{ {\protect\small Absolute value of }$df_{\overline{\protect\mu }%
_{N}}^{(2^{N})}(x)/dx$,{\protect\small \ for }${\protect\small N=1,2,3}$%
{\protect\small \ and }${\protect\small 4}${\protect\small , on a logarithmic
scale as a function of }${\protect\small x}${\protect\small \ in the
interval }${\protect\small 0\leq x\leq 1}${\protect\small . The
proliferation of maxima conveys the development of the hierarchical
structure of repellor preimages. See text. } }
\label{fig.2.3}
\end{figure}

A convenient way to visualize how the preimages for the Feigenbaum attractor
and repellor are distributed and organized is to consider the simpler
arrangements for the preimages of the supercycles' attractors and repellors
of small periods $2^{N}$, $N=1,2,\ldots$ As we have seen, when the period $%
2^{N}$ increases the preimage structures for the attractor and repellor
become more and more involved, with the appearance of new features made up
of an infinite repetition of building blocks. Each of these new blocks is
equivalent to the more dense structures present in the previous $2^{N-1}$
case. In addition all other structures in the earlier $2^{N-2}$, ..., $2^{1}$
cases are still present. Thus a progressively more elaborate organization of preimages is
built upon as $N$ increases, so that the preimage layout for the Feigenbaum
attractor and repellor is obtained as the limiting form of the rank
structure of the fractal boundaries between the finite period attractor
position basins. The fractal boundaries consist of sub-basins of preimages
for the attractor positions separated by preimages of the repellor
positions. As $N$ increases the sizes of these sub-basins decrease while
their numbers increase and the fractal boundaries cover a progressively
larger part of total phase-space. See Figs. \ref{fig.2.3} and \ref{fig.2.6}.

Interestingly, the sizes of all boundary sub-basins vanish in the limit $%
N\rightarrow \infty $, and the preimages of both attractor and repellor
positions become two sets---with dimension equal to the dimension of phase
space---dense in each other. In the limit $N\rightarrow \infty $ there is an
attractor preimage between any two repellor preimages and the other way
round. (The attractor and repellor are two multifractal sets with dimension $%
d_{f}\simeq 0.538...$ \cite{schuster1}.) To visualize this limiting
situation consider that the positions for the repellors and their first
preimages of the $2^{N}$-th supercycle appear located at the inflection
points of $f_{\overline{\mu }_{N}}^{(2^{N})}(x)$, and it is in the close
vicinity of them that the mentioned fractal boundaries form. To illustrate
how the sets of preimage structures for the Feigenbaum attractor and
repellor develop we plot in Fig. \ref{fig.2.3} the absolute value of $\ln \left\vert
df_{\overline{\mu }_{N}}^{(2^{N})}/dx\right\vert $ for $N=1,2,...,4$ vs $x$. 
The maxima in this curve correspond to the inflection points of $f_{%
\overline{\mu }_{N}}^{(2^{N})}(x)$ at which the repellor positions or their
first preimages are located. As shown in Fig. \ref{fig.2.3}, when $N$ increases the
number of maxima proliferate at a rate faster than $2^{N}$.

\subsection{Sequential opening of phase-space gaps}

\begin{figure}[h!]
\centering
\includegraphics[width=.5\textwidth]{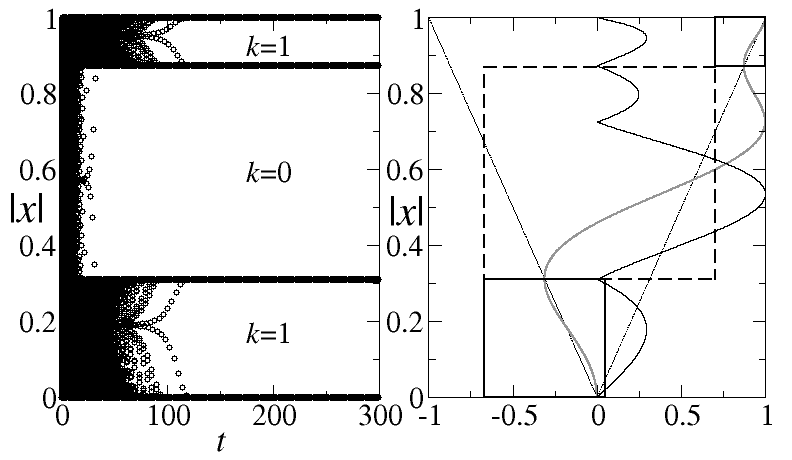}
\caption{ {\protect\small \ Phase-space gap formation for }${\protect\small 
\protect\mu =}\overline{\protect\mu }_{2}${\protect\small . Left panel: time
evolution of a uniform ensemble of $10^4$ trajectories as a function of }$|x|$%
{\protect\small \ (black areas and open circles). The values of the index }$%
{\protect\small k}${\protect\small \ label the order of the gap set. Right
panel: Rotated plots of }$f_{\overline{\protect\mu }_{2}}^{(4)}(x)$%
{\protect\small (gray)\ and }$\left\vert df_{\overline{\protect\mu }%
_{2}}^{(4)}(x)/dx\right\vert ${\protect\small (black) \ vs }$|x|$%
{\protect\small \ as guides for the identification of attractor and repellor
positions.} }
\label{fig.2.4}
\end{figure}

\begin{figure}[h!]
\centering
\includegraphics[width=.5\textwidth]{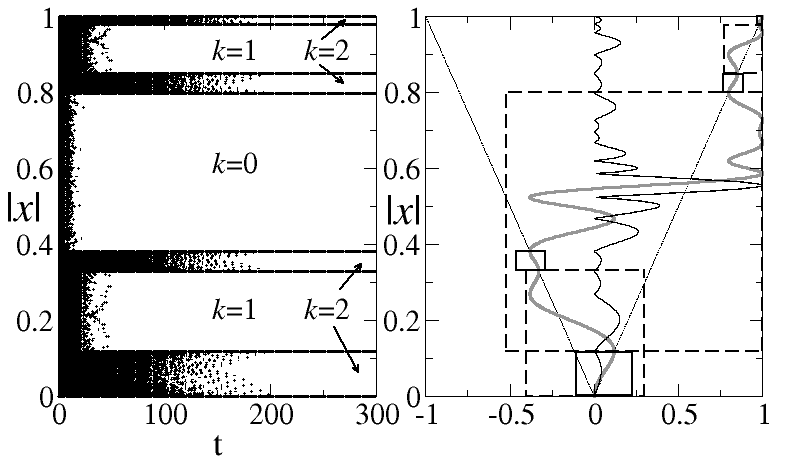}
\caption{ {\protect\small Phase-space gap formation for }${\protect\small 
\protect\mu =}\overline{\protect\mu }_{3}${\protect\small . Left panel: time
evolution of a uniform ensemble of $10^4$ trajectories as a function of }$|x|$%
{\protect\small \ (black areas and open circles). The values of the index }$%
{\protect\small k}${\protect\small \ label the order of the gap set. Right
panel: Rotated plots of }$f_{\overline{\protect\mu }_{3}}^{(8)}(x)$%
{\protect\small (gray) \ and }$\left\vert df_{\overline{\protect\mu }%
_{3}}^{(8)}(x)/dx\right\vert ${\protect\small (black)\ vs }$|x|$%
{\protect\small \ as guides for the identification of attractor and repellor
positions.} }
\label{fig.2.5}
\end{figure}

\begin{figure}[h!]
\centering
\includegraphics[width=.5\textwidth]{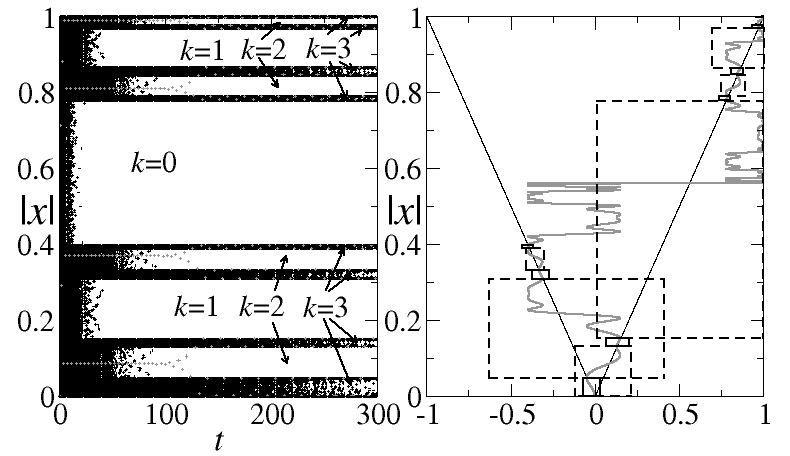}
\caption{ {\protect\small Phase-space gap formation for }${\protect\small 
\protect\mu =\overline{\protect\mu }_{4}}${\protect\small . Left panel: time
evolution of a uniform ensemble of $10^4$ trajectories as a function of }$|x|$%
{\protect\small \ (black areas and open circles). The values of the index }$%
{\protect\small k}${\protect\small \ label the order of the gap set. Right
panel: Rotated plots of }$f_{\overline{\protect\mu }_{4}}^{(16)}(x)$%
{\protect\small (gray)\ and }$\left\vert df_{\overline{\protect\mu }%
_{4}}^{(16)}(x)/dx\right\vert ${\protect\small (black)\ vs }$|x|$%
{\protect\small \ as guides for the identification of attractor and repellor
positions.} }
\label{fig.2.6}
\end{figure}

One way that the preimage structure described above is manifest in the
dynamics is via the successive formation of phase-space gaps that ultimately
give rise to the attractor and repellor multifractal sets. In order to
observe explicitly this process we consider an ensamble of initial
conditions $x_{0}$ spread out uniformly across the interval $-1\leq
x_{0}\leq 1$ and keep track of their positions at subsequent times. In Figs.
\ref{fig.2.4}--\ref{fig.2.6} we illustrate the outcome for the supercycles of periods $2^{2}$, $
2^{3}$, and $2^{4}$, respectively, where we have plotted the time evolution
of an ensemble composed of $10^4$ trajectories. In the left panel of each
figure we show the absolute value of the positions $\left\vert
x_{t}\right\vert $ vs time $t$, while, for comparison purposes, in the right
panel we show the absolute value of $\left\vert x\right\vert $ both \textit{%
vs} $f_{\overline{\mu }_{N}}^{(2^{N})}(x)$ and \textit{vs} $\left\vert df_{%
\overline{\mu }_{N}}^{(2^{N})}/dx\right\vert $ to facilitate identification
of the attractor and repellor positions. The labels $k=0,1,2,\ldots $
indicate the order of the gap set (or equivalently the order of the repellor
generation set). In Fig. \ref{fig.2.4} (with $\mu =\overline{\mu }_{2}$) one observes a
large gap opening first that contains the old repellor ($k=0$) in its middle
region and two smaller gaps opening afterward that contain the two repellors
of second generation ($k=1$) once more around the middle of them. In Fig. \ref{fig.2.5}
(with $\mu =\overline{\mu }_{3}$) we initially observe the opening of a
primary and the two secondary gaps as in the previous $\mu =\overline{\mu }%
_{2}$ case, but subsequently four new smaller gaps open each around the
third generation of repellor positions ($k=2$). In Fig. \ref{fig.2.6} (with $\mu =%
\overline{\mu }_{4}$) we observe the same development as before; however, at
longer times eight additional and yet smaller gaps emerge around each
fourth generation of repellor positions ($k=3$). Naturally, this process
continues indefinitely as $N\rightarrow \infty $ and illustrates the
property mentioned before for $\mu _{\infty }$, that the time evolution at fixed
control parameter values resembles progression from $\mu =0$ up to, in this
section, $\overline{\mu }_{N}$. It is evident in all Figs. \ref{fig.2.4}--\ref{fig.2.6} that
the closer the initial conditions $x_{0}$ are to the repellor positions the
longer times it takes for the resultant trajectories to clear the gap
regions. This intuitively evident feature is essentially linked to the
knowledge we have gained about the fractal boundaries of the preimage
structure, and the observable ``bent over'' portions of these distinct
trajectories in the figures correspond to their passage across the
boundaries. (Since the ensemble used in the numerical experiments is finite
there appear only a few such trajectories in Figs. \ref{fig.2.4}--\ref{fig.2.6}.)

\begin{figure}[h!]
\centering
\includegraphics[width=.5\textwidth]{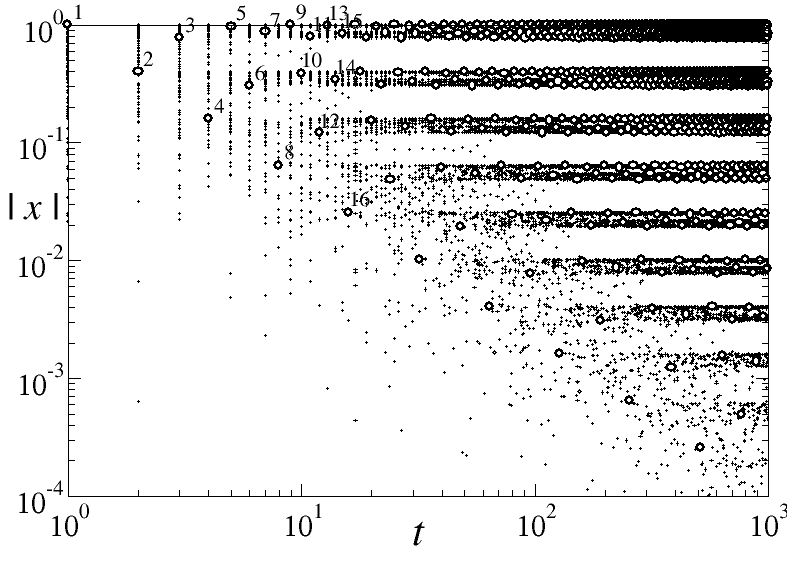}
\caption{ {\protect\small Phase-space gap formation for }$\protect\mu =%
{\protect\small \protect\mu }_{\infty }${\protect\small . The black dots
correspond to time evolution of a uniform ensemble of $10^4$ trajectories as
a function of }$|x|${\protect\small \ vs }${\protect\small t}$%
{\protect\small ,\ both on logarithmic scales. The open circles are the
positions, labeled by the times at which they are reached, for the
trajectory inside the Feigenbaum attractor with initial condition }$%
{\protect\small x}_{{\protect\small 0}}{\protect\small =0}${\protect\small ,
the same as the right panel in Fig. \ref{fig.1.1}.} }
\label{fig.2.7}
\end{figure}

\begin{figure}[h!]
\centering
\includegraphics[width=.5\textwidth]{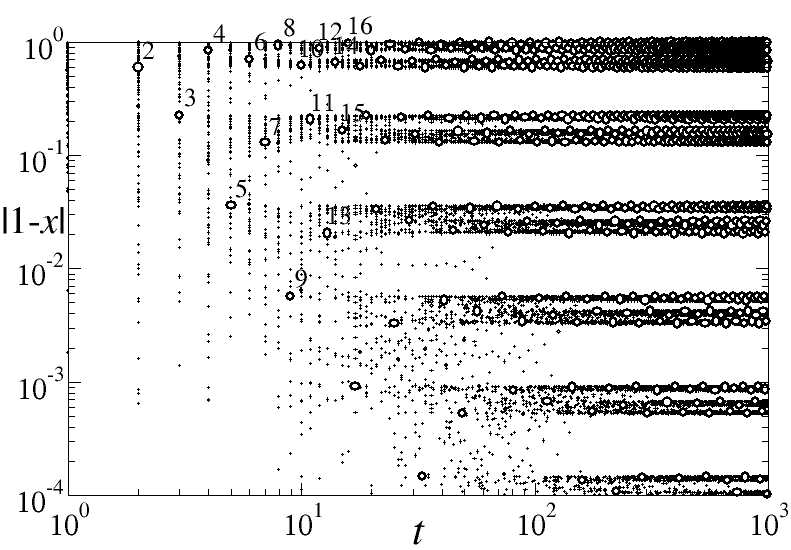}
\caption{ {\protect\small Same as Fig. \ref{fig.2.7} but with replacement of }$|x|$%
{\protect\small \ by }$|{\protect\small 1}-x|${\protect\small . Notice the
change in slope with respect to Fig. \ref{fig.2.7} in the opening of gaps and in the
layout of the positions for the trajectory inside the attractor. } }
\label{fig.2.8}
\end{figure}

To facilitate a visual comparison between the process of gap formation at $%
\mu _{\infty }$ and the dynamics inside the Feigenbaum attractor---as
illustrated by the trajectory in Fig. \ref{fig.1.1} (right panel)---we plot in Fig. \ref{fig.2.7} the time
evolution of the same ensemble composed of $10^4$ trajectories with $\mu
=\mu _{\infty }$. This time we use logarithmic scales for both $\left\vert
x_{t}\right\vert $ and $t$ and then superpose on the evolution of the
ensemble the positions for the trajectory starting at $x_{0}=0$. It is clear
from this figure that the larger gaps that form consecutively all have the
same width in the logarithmic scale of the plot and therefore their actual
widths decrease as a power law, the same power law followed, for instance,
by the position sequence $x_{t}=\alpha ^{-N}$, $t=2^{N}$, $N=0,1,2,\ldots 
$ for the trajectory inside the attractor starting at $x_{0}=0$. This set of
gaps develop in time beginning with the largest one containing the $k=0$
repellor, then followed by a second gap, one of a set of two gaps associated
with the $k=1$ repellor, next a third gap, one gap of a set of four gaps
associated with the $k=2$ repellor, and so forth. The locations of this
specific family of consecutive gaps advance monotonically towards the 
sparsest region of the multifractal attractor located at $x=0$. The remaining
gaps formed at each stage converge, of course, to locations near other
regions of the multifractal, but are not easily seen in Fig. \ref{fig.2.7} because of
the specific way in which this has been plotted (and because of the scale
used). In Fig. \ref{fig.2.8} we plot the same data differently, with the variable $\ln
\left\vert x\right\vert $ replaced by $\ln \left\vert 1-x\right\vert $, where
now another specific family of gaps, one for each value of $k=0,1,2,\ldots $%
, appears, all with the same width on the logarithmic scale; their actual
widths decrease now as $\alpha ^{-2N}$, $N=0,1,2,\ldots $. The locations of
this second family of consecutive gaps advance monotonically towards the
most crowded region of the multifractal attractor located at $x=1$. The time
necessary for the formation of successive gaps of order $k=0,1,2,\ldots $
increases as $2^{k}$ because the duration of equivalent movements of the
trajectories across the corresponding preimage structures involve the $2^{k}$%
-th composed function $f_{\overline{\mu }_{N}}^{(2^{k})}(x)$.

\subsection{Scaling for the rates of convergence to the attractor and repellor%
}

\begin{figure}[h!]
\centering
\includegraphics[width=.5\textwidth]{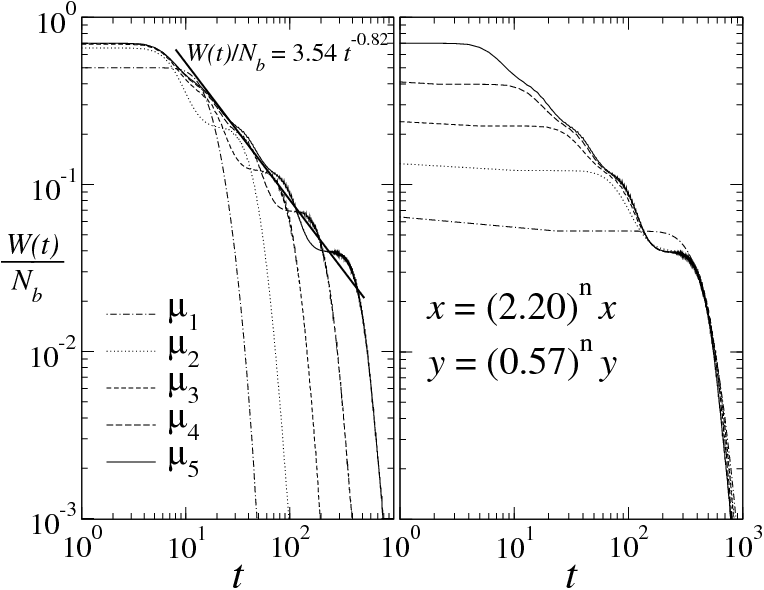}
\caption{ {\protect\small Left panel: Rate }${\protect\small W(t)}$%
{\protect\small , divided by the number of boxes }${\protect\small N}_{b}$%
{\protect\small \ employed,\ of approach to the attractor for the
supercycles of periods }${\protect\small 2}^{{\protect\small N}}$%
{\protect\small , }${\protect\small N=1,2,3,4}${\protect\small \ and }$%
{\protect\small 5}${\protect\small \ on logarithmic scales. The expression
shown corresponds to the power-law decay of the developing logarithmic
oscillations. Right panel: Superposition of the five curves for }$%
{\protect\small W(t)}${\protect\small \ in the left panel via }$%
{\protect\small n}${\protect\small -times repeated rescaling factors shown
for the horizontal }${\protect\small x}${\protect\small \ and vertical }$%
{\protect\small y}${\protect\small \ axes.} }
\label{fig.2.9}
\end{figure}

There is \cite{lyra1} an all-inclusive and uncomplicated way to measure the
rate of convergence of an ensemble of trajectories to the attractor (and to
the repellor) that consists of a single time-dependent quantity. A partition
of phase-space is made of $N_{b}$ equally sized boxes or bins and a uniform
distribution, of $N_{c}$\ initial conditions placed along the interval $%
-1\leq x\leq 1$, is considered again. The number $r$ of trajectories per box
is $r=N_{c}/N_{b}$. The quantity of interest is the number of boxes $W(t)$
that contain trajectories at time $t$. This is shown in Fig. \ref{fig.2.9} on
logarithmic scales for the first five supercycles of periods $2^{1}$ to $%
2^{5}$ where we can observe the following features: In all cases $W(t)$
shows a similar initial nearly constant plateau [$W(t)\simeq \Delta $, $%
1\leq t_{1}\leq t_{0}$, $t_{0}=O(1)$] and a final well-defined decay to
zero. As it can be observed in the left panel of Fig. \ref{fig.2.9} the
duration of the final decay grows (approximately) proportionally to the period 
$2^{N}$ of the supercycle. There is an intermediate slow decay of $W(t)$
that develops as $N$ increases with duration also (just about) proportional
to $2^{N}$. For the shortest period $2^{1}$, there is no intermediate feature
in $W(t)$; this appears first for period $2^{2}$ as a single dip and expands
with one undulation every time $N$ increases by one unit. The expanding
intermediate regime exhibits the development of a power law decay with the
logarithmic oscillations characteristic of discrete scale invariance \cite
{sornette1}. Clearly, the manifestation of discrete invariance is expected
to be associated with the period-doubling cascade. In the right panel of Fig. 
\ref{fig.2.9} we show a superposition of the five curves in Fig. \ref{fig.2.9}
(left panel) obtained via rescaling of both $W(t)$ and $t$ for each curve
according to repeated scale factors.

\begin{figure}[h!]
\centering
\includegraphics[width=.5\textwidth]{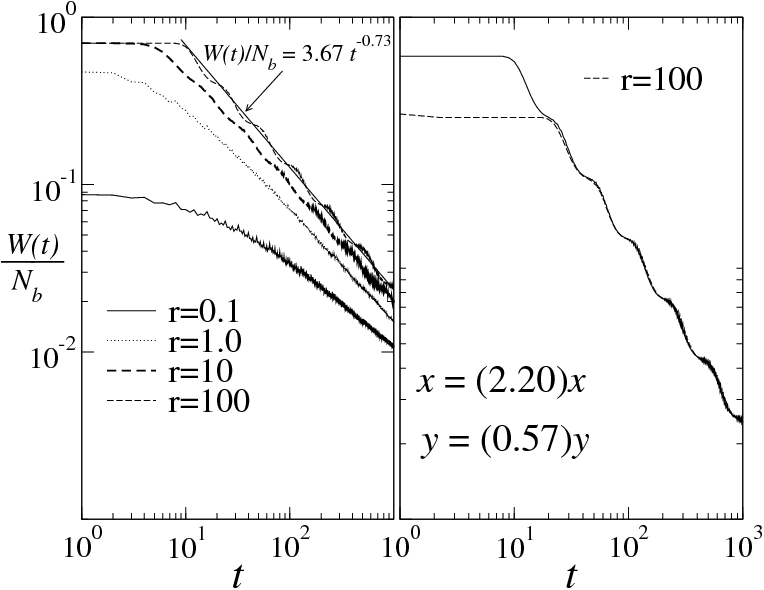}
\caption{ {\protect\small Left panel: Rate }${\protect\small W(t)}$%
{\protect\small \ of approach to the Feigenbaum attractor at }$\protect\mu =%
{\protect\small \protect\mu }_{\infty }${\protect\small \ on logarithmic
scales. The curves correspond to the values given for the number }$%
{\protect\small r}${\protect\small \ of trajectories per box, and the
expression shown corresponds to power law decay with logarithmic
oscillations. Right panel: Superposition of }${\protect\small W(t)}$%
{\protect\small , for }${\protect\small r=100}${\protect\small , on itself
via the rescaling shown (the same as in Fig. \ref{fig.2.9}) for the horizontal }$%
{\protect\small x}${\protect\small \ and vertical }${\protect\small y}$%
{\protect\small \ axes.} }
\label{fig.2.10}
\end{figure}

\begin{figure}[h!]
\centering
\includegraphics[width=.5\textwidth]{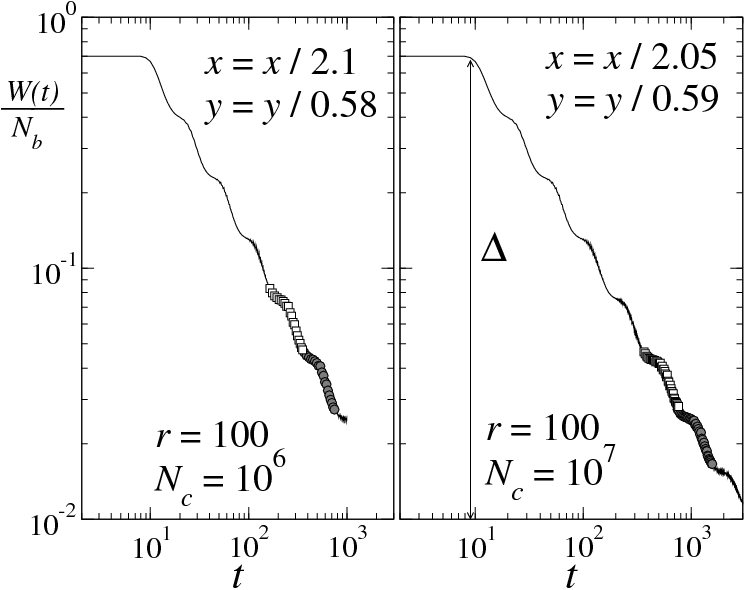}
\caption{ {\protect\small Same as in the right panel of Fig. \ref{fig.2.10} but
obtained with increased precision. In the left panel the number of initial
conditions is }${\protect\small N}_{c}{\protect\small =10}^{6}$%
{\protect\small , while in the right panel }${\protect\small N}_{c}%
{\protect\small =10}^{7}${\protect\small .} {\protect\small The distance} $%
\Delta$ {\protect\small is } $\Delta= (1+|-1/\protect\alpha|)/(1+|-1|)$, 
{\protect\small where }$\protect\alpha${\ is Feigenbaum's constant. This
stems from the fact that all initial conditions out of the interval} $(-1/ 
\protect\alpha, 1)$ {\protect\small take a value inside this interval in the
first iteration.} {\protect\small As can be observed, the scaling factor
for the horizontal axis converges to the exact value }${\protect\small x=2}$%
{\protect\small .} }
\label{fig.2.11}
\end{figure}

The limiting form $W(t)$ for $N\rightarrow \infty $ is shown in the left
panel of Fig. \ref{fig.2.10} for various values of $r$ while in its right
panel we show, for $r=100$, a scale amplification of $W(t)$ with the same
factors employed in Fig. \ref{fig.2.9} for the supercycles with small periods.
The behavior of $W(t)$ at $\mu _{\infty }$ was originally presented in Ref. \cite{lyra1}, 
where the power law exponent $\varphi $ and the logarithmic
oscillation parameter $\Lambda $ in%
\begin{equation}
W(t)\simeq \Delta h\left( \frac{\ln \tau }{\ln \Lambda }\right) \tau
^{-\varphi },\;\tau =t-t_{0},
\label{eq2}
\end{equation}%
were obtained numerically with a precision that corresponds to $r=10$. In
Eq. (\ref{eq2}) $h(x)$ is a periodic function with $h(1)=1$ and $\Lambda $ is the
scaling factor between the periods of two consecutive oscillations. More
recently, in Ref. \cite{grassberger1}, it was pointed out that numerical
estimates of $W(t)$ are subject to large finite-size corrections, and, also,
that $W(t)$ should scale with the intervals in the triadic cantor set
construction of the Feigenbaum attractor \cite{grassberger1}, from which the
value for $\varphi\cong 0.800138194$ is reported. The values for the rescaling
factors in our Figs. \ref{fig.2.9} and \ref{fig.2.10} suffer from these large
finite-size effects due to the relatively small values of $r$ used in the
calculations. This is evident since the time scaling factor obtained from
these data differs by $10\%$ from the exact value of $\Lambda =2$ implied by
the discrete scale invariance property created by the period-doubling
cascade. In Fig. \ref{fig.2.11} we show the rate $W(t)$ and the superposition
of repeated amplifications of itself (as in the right panel of Fig. \ref%
{fig.2.10}) for increasing values of $N_{c}$. We find that the scaling factor $%
\Lambda $ converges to its limit $\Lambda =2$.

\begin{figure}[h]
\centering
\includegraphics[width=.5\textwidth]{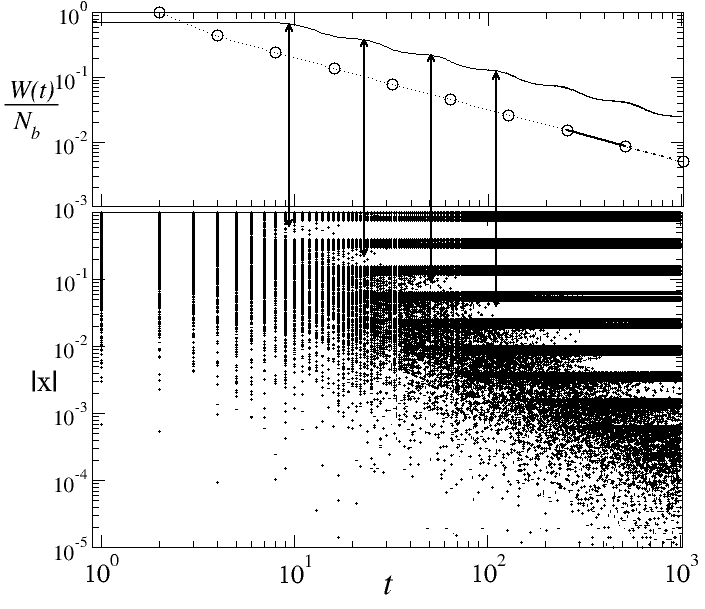}
\caption{ {\protect\small Correspondence between the power law decay with
log-periodic oscillation features of the rate }${\protect\small W(t)}$%
{\protect\small \ with the sequential opening of phase-space gaps. Top
panel: The solid line is }${\protect\small W(t)}${\protect\small \ from Fig.
\ref{fig.2.11} and the open circles values are obtained for }${\protect\small W(t)}$ 
{\protect\small from Eq. (\ref{partition1}) at times }$t={\protect\small 2}^{%
{\protect\small N}}${\protect\small , }$N=1,2,\ldots ${\protect\small \ See
text.} }
\label{fig.2.12}
\end{figure}

We are now in a position to appreciate the dynamical mechanism at work
behind the features of the decay\ rate $W(t)$. From our previous discussion
we know that, every time the period of a supercycle increases from $2^{N-1}$
to $2^{N}$ by a shift the control parameter value from $\overline{\mu }%
_{N-1}$ to $\overline{\mu }_{N}$, the preimage structure advances one stage
of complication in its hierarchy. Along with this, and in relation to the
time evolution of the ensemble of trajectories, an additional set of $2^{N}$
smaller phase-space gaps develops and also a further oscillation takes place
in the corresponding rate $W(t)$ for finite-period attractors. At $\mu =\mu
_{\infty }$ the time evolution tracks the period-doubling cascade progression,
and every time $t$ increases from $2^{N-1}$
to $2^{N}$ the flow of trajectories undergoes equivalent passages across stages 
in the itinerary through the preimage ladder structure, in the development of phase-space gaps, and in
logarithmic oscillations in $W(t)$. In Fig. \ref{fig.2.12} we show the correspondence
between these features quantitatively.

\section{$q$-deformed statistical-mechanical structure}
\label{sec4}
The rate $W(t)$, at the values of time for period doubling, can be obtained
quantitatively from the supercycle diameters $d_{N,m}$. Specifically, $%
W(t)=\Delta \ Z_{\tau }$, $\Delta =(1+\alpha ^{-1})/2$ \cite{note2}, $\tau
=t-t_{0}$, and 
\begin{equation}
Z_{\tau }=\sum_{m=0}^{2^{N-1}-1}d_{N,m},\;\tau =2^{N-1},\;N=1,2,3,...
\label{partition1}
\end{equation}%
Eq. (\ref{partition1})\ is an explicit expression equivalent to the
numerical procedure followed in Ref. \cite{grassberger1} by the use of the
triadic cantor set construction of the Feigenbaum attractor to evaluate the
power law exponent $\varphi $, and from which the value for $\varphi \cong
0.800138194$ is reported. In Fig. \ref{fig.2.12} we have added the results of a
calculation of $W(t)$ at times $\tau =2^{N}$, $N=1,2,\ldots $, according
to Eq. (\ref{partition1}).

The predicted statistical-mechanical structure is exposed when we identify
the (scaled and shifted) decay rate $Z_{\tau }$ as a partition function.
From this viewpoint the diameters $d_{N,m}$ are configurational terms that
in a true statistical-mechanical theory would be expected to obey some
well-defined statistical weights, such as Boltzmann factors. To check on
this we proceed to determine their time dependence given by the bifurcation
index $N$. The $d_{N,m}$ scale (asymptotically) with $N$ for $m/2^N$ fixed as%
\begin{equation}
d_{N,m}\simeq \alpha _{y}^{-N+1},\;N \;large  \label{diameters2}
\end{equation}%
where the $\alpha _{y}$ are universal constants obtained, for instance, from
the finite jump discontinuities of Feigenbaum's trajectory scaling function $%
\sigma (y)=\lim_{n\rightarrow \infty }d_{N,m+1}/d_{N,m}$, $%
y=\lim_{N\rightarrow \infty }m/2^{N}$ \cite{schuster1}. The two largest
discontinuities of $\sigma (y)$\ correspond to the thinner and the fuller
regions of the multifractal attractor, and for these two regions we have,
respectively, $d_{N,0}\simeq \alpha ^{-N+1}$ and $d_{N,1}\simeq \alpha
^{-2(N-1)}$. [The length of the first diameter is $d_{1,0}=1$ and the
equality in Eq. (\ref{diameters2}) is approached rapidly with increasing $N$%
.] The power law in Eq. (\ref{diameters2}) can be rewritten as a $q$%
-exponential ($\exp _{q}(x)\equiv \lbrack 1-(q-1)x]^{-1/(q-1)}$) via use of
the identity $A^{-N+1}\equiv (1+\beta )^{-\ln A/\ln 2}$, $\beta =2^{N-1}-1$,
that is,%
\begin{equation}
d_{N,m}\simeq \exp _{q_{y}}(-\beta \nu _{y}),  \label{diameters3}
\end{equation}%
where $q_{y}=1+\nu _{y}^{-1}$, $\nu _{y}=\ln \alpha _{y}/\ln 2$, and $\beta
=\tau -1=2^{N-1}-1$. Similarly, the partition function $Z_{\tau }\simeq \tau
^{-\varphi }$ (or $Z_{\tau }\simeq \varepsilon ^{-N+1}$), with $\varphi =\ln
\varepsilon /\ln 2$ and $\tau =2^{N-1}$, can be expressed as%
\begin{equation}
Z_{\tau }\simeq \exp _{Q}(-\beta \varphi ),  \label{partition2}
\end{equation}%
where $Q=1+\varphi ^{-1}$ and once more $\beta =\tau -1=2^{N-1}-1$.

Our main contention becomes evident when Eqs. (\ref{diameters3}) and (\ref%
{partition2}) are used in Eq. (\ref{partition1}), to yield%
\begin{equation}
\exp _{Q}(-\beta \varphi )\simeq \sum_{y}\exp _{q_{y}}(-\beta \nu _{y}).
\label{partition3}
\end{equation}%
Eq. (\ref{partition3}) is similar to a basic statistical-mechanical
expression; the quantities in it take the following parts: $\beta $ an
inverse temperature, $\varphi $ a free energy (or the product $s=-\beta
\varphi $ a Massieu thermodynamic potential, or entropy), and the $\nu _{y}$
configurational energies. However, the equality involves $q$-deformed
exponentials in place of ordinary exponential functions that would be
recovered when $Q=q_{m}=1$.\ It is worth noticing that there is a
multiplicity of $q$-indices associated to the configurational weights in Eq.
(\ref{partition3}); however, their values form a well-defined family \cite%
{robledo3} determined by the discontinuities of Feigenbaum's function $%
\sigma $.

\begin{figure}[h!]
\centering
\includegraphics[width=.5\textwidth]{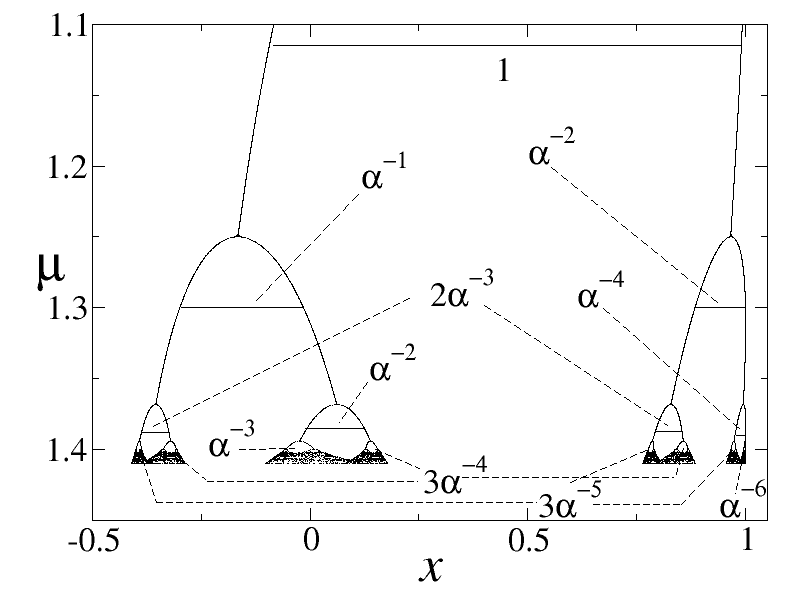}
\caption{
{\small Sector of the bifurcation tree for the logistic map }$%
{\small f}_{\mu }{\small (x)}${\small \ that shows the formation of a Pascal
triangle of diameter lengths according to the scaling approximation
explained in the text, where }${\small \alpha \simeq 2.5091}${\small \ is the
pertinent universal constant. }
}
\label{fig.2.13}
\end{figure}

In the spirit of a mean field approximation we suppose that, for a given
value of $N$, e.g., $N=3$, the diameters $d_{N,m}$ that are of similar
lengths have actually equal length, and this length is obtained from those of
the shortest or longest diameters via a simple scale factor; e.g., $%
d_{3,3}=d_{3,2}=$ $\alpha ^{-1}d_{3,0}=$ $\alpha d_{3,1}$. This initiates a
degree of degeneracy in the lengths that subsequently spreads all the way
through the bifurcation tree. See Fig. \ref{fig.2.13} \cite{linage1}. As a result of
this approximation, the $d_{N,m}$ scale with increasing $N$ according to a
binomial combination of the scale factors of those diameters that converge
to the most crowded and most sparse regions of the multifractal attractor.
To be precise, the $2^{N-1}$ diameters at the $N$-th supercycle have lengths
equal to $\alpha ^{-(N-1-l)}\alpha ^{-2k}$ and occur with multiplicities $%
{N-1 \choose k}$, where $k=0,1,...,N-1$. As shown in Fig. \ref{fig.2.13} the diameters
create a Pascal triangle across the bifurcation cascade. This feature
greatly simplifies the evaluation of the partition function and directly
yields

\begin{equation}
Z_{\tau }=\sum_{k=0}^{N-1}{N-1 \choose k}\alpha ^{-(N-1-k)}\alpha
^{-2k}=\left( \alpha ^{-1}+\alpha ^{-2}\right) ^{N-1},  \label{partition4}
\end{equation}%
$\tau =2^{N-1}$. We obtain $\varepsilon =\left( \alpha ^{-2}+\alpha
^{-1}\right) =1.7883$, $\varphi =0.8386$, and $Q=2.1924$, a surprisingly
good approximation when compared to the numerical estimates $\varphi =0.8001$%
 and $Q=2.2498$ of the exact values. Under this approximation all the
indices $q_{y}$ in Eq. (\ref{partition3}) are equal, $q_{y}=q=1+\nu ^{-1}$, $%
\nu =\ln \alpha /\ln 2$, and Eq. (\ref{partition3}) becomes%
\begin{equation}
\exp _{Q}(-\beta \varphi )=\sum_{k=0}^{N-1}\Omega (N-1,k)\exp _{q}(-\beta
\nu ),  \label{partition5}
\end{equation}%
where $\Omega (N-1,k)=\alpha ^{-l}$ ${N-1 \choose k}$. In thermodynamic
language, the approach to the attractor described by Eq. (\ref{partition3})
or (\ref{partition5}) is a cooling process $\beta \rightarrow \infty $ in
which the free energy (or energy) $\varphi $ is fixed and therefore the
entropy $s=-\beta \varphi $ is linear in $\beta $. It is instructive to
define an ``energy landscape'' for the Feigenbaum attractor as being composed
of an infinite number of ``valleys'' whose equal-valued minima at $\beta
\rightarrow \infty $ coincide with the points of the attractor on the
interval $[-\alpha ^{-1},1]$ \cite{note2}. When $\beta =2^{N-1}-1$, $N$
finite, the valleys merge into $2^{N-1}$ intervals of widths equal to the
diameters $d_{N,m}$.

As established some time ago, the so-called thermodynamic formalism \cite{beck1}
is built around the statistical-mechanical framework followed by the
geometric properties of multifractals. The partition function formulated to
study their properties, like the spectrum of singularities $f(\widetilde{%
\alpha })$ \cite{beck1}, is 
\begin{equation}
Z(\widetilde{\mathsf{\tau }},\mathsf{q})\equiv \sum_{m}^{M}p_{m}^{\widetilde{%
\mathsf{\tau }}}l_{m}^{-\mathsf{q}},  \label{partition6}
\end{equation}%
where the $l_{m}$ (in one-dimensional systems) are $M$ disjoint interval
lengths that cover the multifractal set and the $p_{m}$ are probabilities
given to these intervals. The standard practice consists of demanding that $%
Z(\mathsf{\tau },\mathsf{q})$ neither vanishes nor diverges in the limit $%
l_{m}\rightarrow 0$ for all $m$ (and consequently $M\rightarrow \infty $).
Under this condition the exponents $\widetilde{\mathsf{\tau }}$ and $\mathsf{%
q}$ define a function $\widetilde{\mathsf{\tau }}(\mathsf{q})$ from which $f(%
\widetilde{\alpha })$ is obtained via Legendre transformation \cite{beck1}.
When the multifractal is an attractor its elements are ordered dynamically,
and for the Feigenbaum attractor the trajectory with initial condition $%
x_{0}=0$ generates in succession the positions that form the diameters,
generating all diameters $d_{N,m}$ for $N$ fixed between times $\tau
=2^{N-1} $ and $\tau =3\times 2^{N-1}$. Because the diameters cover the
attractor it is natural to choose the covering lengths at stage $N$ to be $%
l_{m}^{(N)}=$ $d_{N,m}$ and to assign to each of them the same probability $%
p_{m}^{(N)}=1/2$. For example, within the two-scale approximation to the
Feigenbaum multifractal \cite{beck1}, $l_{k}^{(N)}=\alpha ^{-(N-1-k)}\alpha
^{-2k}$, the condition $Z(\widetilde{\mathsf{\tau }},\mathsf{q})=1$
reproduces Eq. (\ref{partition4}) when $p_{m}^{(N)}=$ $\tau ^{-1}=2^{-N+1}$,
with $\widetilde{\mathsf{\tau }}=1$ and $\mathsf{q}=-\varphi $. It should be
kept in mind that the ``static'' partition function $Z(\widetilde{\mathsf{\tau 
}},\mathsf{q})$ is not meant to distinguish between chaotic and critical
(vanishing $\lambda $) multifractal attractors as we do here. As we
emphasize below, it is the functional form of the link between the
probabilities $p_{m}^{(N)}$ and actual time $\tau $ that determines the
nature of the statistical mechanical structure of the dynamical system.

A crossover to $q=1$ ordinary statistics when a critical attractor turns
chaotic; this can be explained as follows. We first recall that multifractal sets
and their statistical-mechanical properties can be retrieved by means of the
recursive method of backward iteration of chaotic maps \cite{mccauley1}. A
chaotic unimodal map has a two-valued inverse and given a position $x=x_{n}$
a binary tree is formed under backward iteration, so there are $2^{n}$
initial conditions $x_{0}$ for trajectories that lead to $x_{n}$. Since in
this case the Lyapunov exponent is positive $\lambda >0$, lengths expand
under forward iteration according to $l\sim \exp (\lambda n)$ and contract
under backward iteration as $l\sim \exp (-\lambda n)$. We can define, as
above, a set of covering lengths $D_{n,m}=$ $\Delta _{m}\exp (-\lambda n)$,
where $m$ relates to the initial condition $x_{0}$ and use of them in a
partition function like that in Eq. (\ref{partition1}) gives%
\begin{equation}
\exp (-\beta \varphi )=\sum_{m}\Delta _{m}\exp (-\beta \lambda ),
\label{partition7}
\end{equation}%
where now $\beta =n$. Keeping in mind Pesin's theorem $\varphi $ is plainly
identified as the Kolmogorov-Sinai entropy. Now, the crossover from $q$%
-deformed statistics to ordinary $q=1$ statistics can be observed for
control parameter values in the vicinity of the Feigenbaum attractor, $\mu
\gtrsim \mu _{\infty }$, when the attractor consists of $2^{\overline{n}}$
bands, $\overline{n}$ large. The Lyapunov coefficient $\lambda $ of the
chaotic attractor decreases with $\Delta \mu =\mu -$ $\mu _{\infty }$ as $%
\lambda \varpropto 2^{-\overline{n}}\sim \Delta \mu ^{\kappa }$, $\kappa
=\ln 2/\ln \delta _{F}$, where $\delta _{F}$ is the Feigenbaum
constant that measures the rate of development of the bifurcation tree in
control parameter space \cite{schuster1}. The chaotic orbit consists of an
interband periodic motion of period $2^{\overline{n}}$ and an intraband
chaotic motion. The expansion rate $\sum_{i=0}^{\tau -1}\ln \left\vert
df_{\mu }(x_{i})/dx_{i}\right\vert $ fluctuates with increasing amplitude as 
$\ln \tau $ for $\tau <2^{\overline{n}}$ but converges to a fixed number
that grows linearly with $\tau $ for $\tau \gg 2^{\overline{n}}$ \cite{mori1}%
. This translates as dynamics with $q\neq 1$ for $\tau <2^{\overline{n}}$
but ordinary dynamics with $q=1$ for $\tau \gg 2^{\overline{n}}$.


\section{Summary}
\label{sec5}
As stated, the dynamics of critical attractors in low-dimensional nonlinear
maps is a suitable phenomenon for assessing the limits of validity and
generalizations of ordinary statistical mechanics. There are now positive
indications that the multifractal critical attractors present in these maps
play this role, since, as it turns out, the two sets of dynamical properties---
inside and towards the Feigenbaum attractor---appear combined in a $q$%
-deformed statistical-mechanical structure \cite{robledo00}. To obtain this
remarkable property it is necessary to have access to detailed information
for these two different types of properties. The dynamics \textit{at} the
attractor (for both trajectories and sensitivity to initial conditions) has
been analyzed in detail before for period doubling \cite{robledo2,robledo3} 
and for the quasiperiodic \cite{robledo7} transition to chaos.
But the dynamics \textit{on the way to} the attractor is only now offered 
as far as we know.

To begin with, we studied the properties of the first few members of the
family of superstable attractors of unimodal maps with quadratic maxima and
obtained a precise understanding of the complex labyrinthine dynamics that
develops as their period $2^{N}$ increases. The study is based on the
determination of the function $t_{f}(x_{0})$, the time of flight for a
trajectory with initial condition $x_{0}$ to reach the attractor or
repellor. The function $%
t_{f}(x_{0})$ was determined for \textit{all} initial conditions $x_{0}$ in
a partition of the total phase-space $-1\leq x_{0}\leq 1$, and this provides
a complete picture for each attractor-repellor pair. We observed how the
fractal features of the boundaries between the basins of attraction of the
positions of the periodic orbits develop a structure with hierarchy, and how
this in turn reflects on the properties of the trajectories. The set of
trajectories produces an ordered flow towards the attractor or towards the
repellor that reflects the ladder structure of the sub-basins that constitute
the mentioned boundaries. As $2^{N}$ increases there is sensitivity to the
final position for almost all $x_{0}$, and there is a transient
exponentially-increasing sensitivity to initial conditions for almost all $%
x_{0}$. We observed that transient chaos is the manifestation of the
trajectories' controlled flow out of the fractal boundaries, which suggests
that for large $2^{N}$ the flow becomes an approximately self-similar
sequence of stages. As a final point, in the Appendix, we look at the
closing segment of trajectories at which a very fast convergence to the
attractor positions occurs. We found ``universality class'' features, as the
trajectories and sensitivity to initial conditions are replicated by a RG
fixed-point map obtained under functional composition and rescaling. This
map has the same $q$-deformed exponential closed form found to hold also for
the pitchfork and tangent bifurcations of unimodal maps \cite{robledo5,robledo6}.

Subsequently, we examined the process followed by an ensemble of
uniformly distributed initial conditions $x_{0}$ across the phase-space to
arrive at the Feigenbaum attractor, or get captured by its corresponding
repellor. Significantly, we gained understanding concerning the
dynamical ordering in $x_{0}$, in relation to the construction of the
families of phase-space gaps that support the attractor and repellor, and
about the rate of approach of trajectories towards these multifractal sets,
as measured by the fraction of bins $W(t)$ still occupied by trajectories at
time $t$. An important factor in obtaining this knowledge has been the
consideration of the equivalent dynamical properties for the supercycles of
small periods in the bifurcation cascade. As we have seen, a doubling of the
period introduces well-defined additional elements in the hierarchy of the
preimage structure, in the family of phase-space gaps, and in the
log-periodic power law decay of the rate $W(t)$. We have then corroborated
the wide-ranging correlation between time evolution at $\mu _{\infty }$ from 
$t=0$ up to $t\rightarrow \infty $ with the static period-doubling cascade
progression from $\mu =0$ up to $\mu _{\infty }$. As a result of this we
have acquired an objective insight into the complex dynamical phenomena that
fix the decay\ rate $W(t)$.\ We have clarified the genuine mechanism by
means of which the discrete scale invariance implied by the log-periodic
property in $W(t)$ arises, that is, we have seen how its self-similarity
originates in the infinite hierarchy formed by the preimage structure of the
attractor and repellor. The rate $W(t)$ can be obtained quantitatively [see
Eq. (\ref{partition1})] from the supercycle diameters $d_{N,m}$. These basic data
descriptive of the period-doubling route to chaos are also a sufficient
ingredient in the determination of the anomalous sensitivity to initial
conditions for the dynamics inside the Feigenbaum attractor \cite{robledo3}.

Finally, the case is made that there is a statistical-mechanical property
underlying the dynamics of an ensemble of trajectories en route to the
Feigenbaum attractor (and repellor). Eq. (\ref{partition1}) is identified as a partition
function built of $q$-exponential weighted configurations, and in turn, the
fraction $Z_{\tau }$ of phase-space still occupied at time $\tau $ is seen
to have the form of the $q$-exponential of a thermodynamic potential
function. This is argued to be a concrete, clear and genuine manifestation
of $q$-deformation of ordinary statistical mechanics where arguments can be
made explicit and rigorous. There is a close resemblance with the
thermodynamic formalism for multifractal sets, but it should be stressed
that the deviation from the usual exponential statistics is dynamical in
origin, and due to the vanishing of the (only) Lyapunov exponent.
\newline

\textbf{Acknowledgments.} We are grateful to Dan Silva for useful
preliminary studies, to Jorge Velazquez Castro for valuable help and to Hugo
Hern\'{a}ndez Salda\~{n}a for interesting discussions. Partial support by
DGAPA-UNAM and CONACyT (Mexican agencies) as well as SIMUMAT (Comunidad de Madrid, 
Spanish agency) is acknowledged.

\section{Appendix: Super-strong insensitivity to initial conditions}

The ordinary Lyapunov exponent $\lambda \equiv \lim_{t\rightarrow \infty
}t^{-1}\ln \left. df_{\overline{\mu }_{N}}^{(t)}(x)/dx\right\vert _{x=0}$
for the supercycle attractors diverges to minus infinity [$df_{\overline{\mu 
}_{N}}^{(t)}(x)/dx=0$ at $x=0$], therefore the sensitivity to initial
conditions $\xi _{t}$ cannot have an exponential form $\xi _{t}=\exp
(\lambda t)$ with $\lambda <0$. This appendix is a brief account of the
determination of $\xi _{t}$ at the closing point of approach to the
supercycle attractors.
\begin{figure}[h!]
\centering
\includegraphics[width=.5\textwidth]{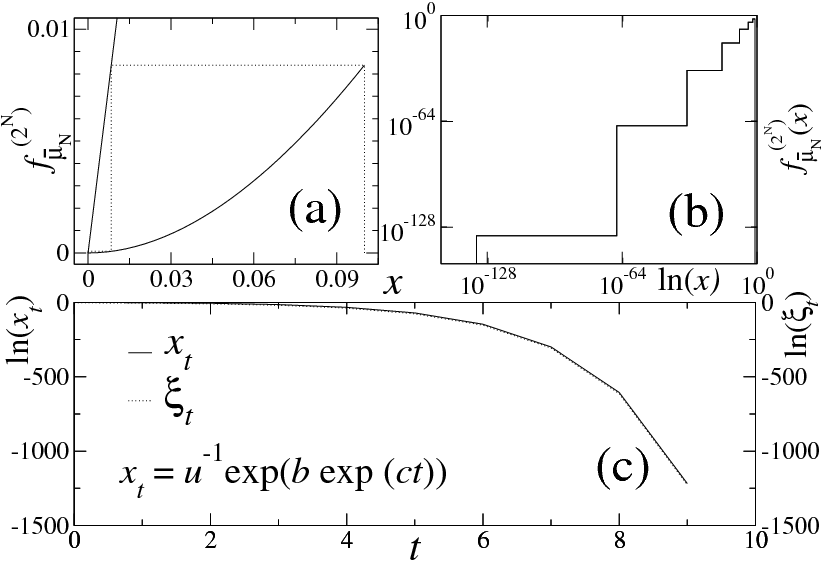}
\caption{ {\protect\small Panel (a): Detail of a trajectory in its final
stage of approach to the }${\protect\small x=0}${\protect\small \ attractor
position (dotted line) for the map }$f_{\overline{\protect\mu }%
_{N}}^{(2^{N})}(x)${\protect\small \ (solid line). In this example }$%
{\protect\small N=3}${\protect\small . Panel (b): Same as (a) in double
logarithmic scale. Panel (c): Trajectory }$x_{t}${\protect\small \ and
sensitivity to initial conditions }$\protect\xi _{t}${\protect\small \ in
logarithmic scale versus time }${\protect\small t}${\protect\small . Both
functions are indistinguishable. Note that there is an ultra-rapid
convergence, of only a few time steps, to the origin }${\protect\small x=0}$%
{\protect\small . See text.} }
\label{fig.1.13}
\end{figure}

Representative results for the last segment of a trajectory and the
corresponding sensitivity $\xi _{t}$ obtained from a numerical investigation
are shown in Fig. \ref{fig.1.13}.\ Only the first two steps of a trajectory
with $x_{0}=0.1$ of the map $f_{\overline{\mu }_{3}}^{(2^{3})}$ can be seen
in Fig. \ref{fig.1.13}(a). A considerable enlargement of the spatial scale
(which requires computations of extreme precision \cite{mapm1}) makes it
possible to observe a total of seven steps ($56$ iterations in the original
map), as shown with the help of logarithmic scales in Fig. \ref{fig.1.13}(b).
Fig. \ref{fig.1.13}(c) shows both the same trajectory and the sensitivity $%
\xi _{t}\equiv dx_{t}/dx_{0}$ in a logarithmic scale for $x_{t}$ and $\xi
_{t}$ and a normal scale for the time $t$. The trajectory is accurately
reproduced [indistinguishable from the curve in Fig. \ref{fig.1.13} (c)] by
the expression $x_{t}=u^{-1}\exp (b\exp ct)$, $b=\ln ux_{0}$ (with $x_{0}>0$%
), and $c=\ln 2$, where $u>0$ is obtained from the form $|f_{\overline{\mu }%
_{N}}^{(2^{N})}|\simeq ux^{2}$ taken by the $2^{N}$-th composed map close
to $x=0$. This expression for\ $x_{t}$ is just another form of writing $%
ux_{t}=(ux_{0})^{2^{t}}$, the result of repeated iteration of $ux^{2}$. For
the logarithm of the sensitivity we have $\ln \xi _{t}=-\ln x_{0}+t\ln 2+\ln
x_{t}$ where the last (large negative) term dominates the first two. Thus,
we find that the sensitivity decreases more quickly than an exponential, and, more
precisely, decreases as the exponential of an exponential.

We note that, associated with the general form $f_{\overline{\mu }%
_{N}}^{(2^{N})}(x)\simeq ux^{2}$ of the map in the neighborhood of $x=0$,
there is a map $f^{\ast }(x)$ that satisfies the functional composition and
rescaling equation $f^{\ast }(f^{\ast }(x))=a^{-1}f^{\ast }(ax)$ for some
finite value of $a$ and such that $f^{\ast }(x)=ux^{2}+o(x^{4})$. The
fixed-point map $f^{\ast }(x)$ possesses properties common to all
superstable attractors of unimodal maps with a quadratic extremum. Indeed,
there is a closed form expression that satisfies these conditions,  
which is $f^{\ast }(x)=x\exp _{q}(u^{q-1}x).$ 
The fixed-point map equation is satisfied with 
$a=2^{1/(q-1)}$ and $q=1/2$. The same type of RG solution has been
previously found to exist for the tangent and pitchfork bifurcations of
unimodal maps with general nonlinearity $z>1$ \cite{hu1,robledo5,robledo6}. 
Use of the map $f^{\ast }(x)$ reproduces the trajectory $%
x_{t}/x_{0}$ and sensitivity $\xi _{t}$ shown in Fig. \ref{fig.1.13} (c).


\begin{thebibliography}{99}
\bibitem{note1} Here we use the term $q$-deformed statistical mechanics with
reference to the occurrence of statistical-mechanical expressions where
instead of the ordinary exponential and logarithmic functions appear their $%
q $-deformed counterparts. The $q$-deformed exponential function is defined
as $\exp _{q}(x)\equiv \lbrack 1+(1-q)x]^{1/(1-q)}$, and its inverse, the $q$%
-deformed logarithmic function as $\ln _{q}y\equiv (y^{1-q}-1)/(1-q)$. The
ordinary exponential and logarithmic functions are recovered when $q=1$.
\end{thebibliography}
\end{document}